\documentclass[aps, reprint, longbibliograhy, superscriptaddress]{revtex4-2}
\usepackage{graphicx}
\usepackage{amsmath, amssymb}
\usepackage{bm}
\usepackage[version=4]{mhchem}
\usepackage[colorlinks=true, linkcolor=blue, citecolor=blue, urlcolor=blue]{hyperref}
\usepackage{siunitx}
\DeclareSIUnit{\rydberg}{Ryd}
\usepackage{makecell}
\makeatletter
\def\@hangfrom@section#1#2#3{\@hangfrom{#1#2}#3}%\MakeTextUppercase{#3}}%
\def\@hangfroms@section#1#2{#1#2}%\MakeTextUppercase{#2}}%
\makeatother

\begin{document}

\title[]{Anharmonic lattice dynamics study of phonon transport in layered and molecular-crystal indium iodides}% Force line breaks with \\

\author{Takuma Shiga}
\email{shiga@toyota-ti.ac.jp}
\affiliation{
Mechanical Material Engineering Laboratory, Toyota Technological Institute, Nagoya, Aichi 468-8511, Japan
}

\author{Yoshikazu Mizuguchi}
\affiliation{
Department of Physics, Tokyo Metropolitan University, Hachioji, Tokyo 192-0397, Japan
}

\author{Hiroshi Fujihisa}
\affiliation{
National Metrology Institute of Japan (NMIJ), National Institute of Advanced Industrial Science and Technology (AIST), Tsukuba, Ibaraki 305-8565, Japan
}

\date{\today}

\begin{abstract}
Indium iodides, which adopt layered or molecular-crystal-like arrangements depending on composition, are expected to exhibit low lattice thermal conductivity because of their heavy constituent atoms and weak \ce{In}--\ce{I} bonding. In this work, we employed first-principles anharmonic lattice dynamics calculations to systematically investigate phonon transport in indium iodides from particle- and wave-like perspectives. The calculated lattice thermal conductivities of both materials remained below \SI{1}{\watt \per \meter \per \kelvin} over a broad temperature range. Notably, the influence of wave-like phonon transport differed by composition: in \ce{InI3}, the wave-like contribution became comparable to the particle-like Peierls contribution, whereas it remained negligible in \ce{InI}. We also investigated the thermal transport properties of the experimentally reported high-pressure phase of \ce{InI3}. Motivated by experimental indications of stacking faults and partial disorder in indium site occupancy within the rhombohedral phase, we constructed several ordered structural models with different stacking sequences. These stacking sequences exhibited no significant energetic preference and had similar lattice thermal conductivities, suggesting that in-plane thermal transport is largely governed by the vibrational properties of the \ce{In2I6} layers themselves rather than by the specific stacking sequence. These findings provide insight into phonon transport in layered and molecular-crystal systems with structural complexity and contribute to a broader understanding of thermal transport mechanisms in layered and molecular-crystal-like materials.
\end{abstract}

\maketitle

\section{\label{sec:intro}Introduction}
Layered compounds are known to exhibit unique electrical, optical, and thermal properties due to their inherent bonding characteristics and structural anisotropies \cite{RN4493, RN4494, RN4495, RN4502}. Beyond monolayer forms, structural modifications, such as moir\'{e} patterns, heterostructures, and Janus configurations, can induce distinct chemical and physical functionalities \cite{RN4492, RN4496, RN4497, RN4498}. In recent years, the exploration of layered and low-dimensional materials for thermal management (e.g., insulation, dissipation, and energy conversion) has intensified, often supported by data-driven approaches using machine learning \cite{RN4501, RN4503, RN4504, RN4505, RN4468}. However, because of the diverse bonding environments in these systems, many layered materials display polymorphism and structural variability, including different crystal symmetries and stacking orders \cite{RN4500}. Consequently, material predictions based solely on physical descriptors, interatomic potential models, or machine learning methods trained on specific structures may have limited transferability across a wide range of materials. Therefore, alongside these explorations, detailed characterization and systematic property evaluation of individual materials—particularly those with complex structures—are essential.

Among layered materials composed of indium and iodine, orthorhombic \ce{InI} \cite{RN4490, RN4491} and monoclinic \ce{InI3} based on \ce{In2I6} dimers \cite{RN3645} have been studied for decades \cite{RN3788, RN4487}. Notably, InI possesses a high average atomic number and a relatively large band gap \cite{RN3824}, making it a promising candidate for applications such as X-ray and gamma-ray detectors \cite{RN4488, RN4489}, as well as infrared optical devices \cite{RN4499}. Both \ce{InI} and \ce{InI3}  are expected to exhibit ultralow thermal conductivity because of their large atomic masses and weak interlayer or intermolecular interactions. Although previous experimental and theoretical studies have partially elucidated the thermophysical properties of \ce{InI} and \ce{InI3} \cite{RN3597, RN3595, RN4032, RN4033, RN3780, RN3787}, comprehensive investigations of their thermal transport properties from a microscopic phonon perspective remain limited.

Recently, newly identified crystalline phases obtained through high-pressure treatment or synthesis have attracted considerable interest. Under high-pressure conditions, structural transformations often occur alongside changes in coordination number, and the resulting modifications in physical properties are of particular interest as well \cite{RN4475, RN4476, RN4477, RN4478, RN4479}. For example, a phase transformation to a metallic $P4/nmm$ structure has been theoretically predicted for InI at pressures of $\sim$\SI{17}{\giga \pascal} \cite{RN4472}. Although such coordination-driven transformations are noteworthy, layered compounds also commonly undergo pressure-induced layer slippage, which can lead to changes in stacking sequences \cite{RN4480, RN4481, RN4483, RN4484, RN4485}. Indeed, a recent experimental study on \ce{InI3} demonstrated that mechanical pressure induces a structural transformation from the monoclinic low-pressure phase to a rhombohedral high-pressure phase composed of stacked, edge-sharing \ce{In2I6} octahedral layers \cite{RN3593}.

In this study, we investigate the thermal transport properties of \ce{InI} and \ce{InI3} using first-principles-based anharmonic lattice dynamics. For high-pressure \ce{InI3}, although a rhombohedral structure has been reported, it exhibits numerous stacking faults, and its precise atomic configuration remains unclear. To address this uncertainty, we construct several structural models of the high-pressure phase by considering different stacking arrangements of \ce{In2I6} layers. We then systematically evaluate their relative energetic stabilities and corresponding thermal transport properties.

\section{\label{sec:method}Computational Details}
\begin{figure}[t!]
	\centering
	\includegraphics[width=0.45\textwidth]{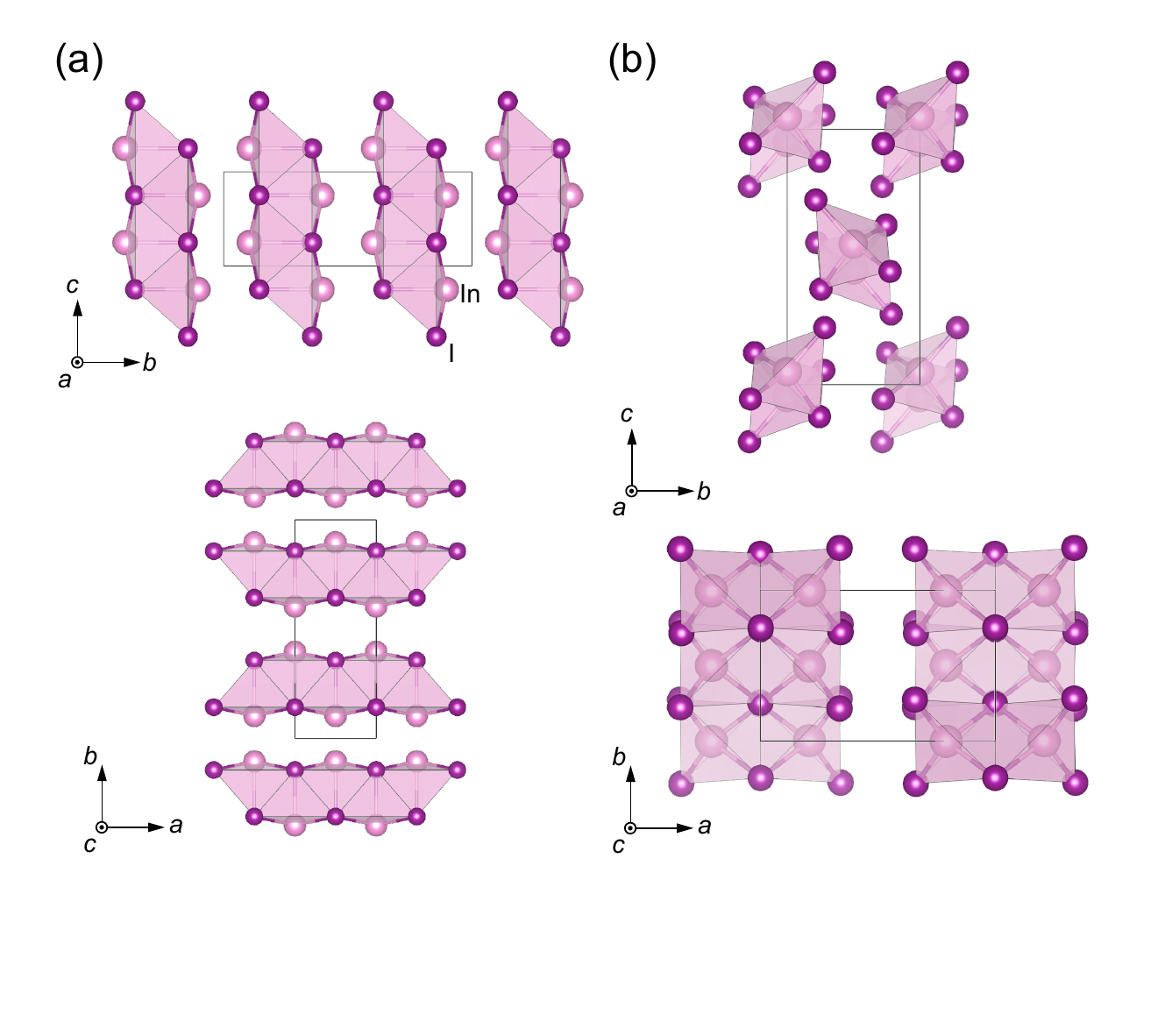}
	\caption{
	Schematic illustration of (a) \ce{InI} and (b) low-pressure \ce{InI3} structures, visualized using the \texttt{VESTA} software \cite{RN3640}.
	}
	\label{fig1}
\end{figure}

To structurally optimize \ce{InI} and \ce{InI3}, we performed density functional theory (DFT) calculations using the \texttt{Quantum ESPRESSO} package \cite{RN372, RN3231}. Following previous studies, we adopted the orthorhombic $Cmcm$ structure (No. 63, $Z = 4$) for \ce{InI} \cite{RN4490, RN4491} and the monoclinic $P2_1/c$ (No. 14, $Z = 4$) for low-pressure \ce{InI3} \cite{RN3645}, as illustrated in Fig. \ref{fig1}. The calculations employed the generalized gradient approximation functional parameterized by Perdew--Burke--Ernzerhof \cite{RN3259} and the projector-augmented wave method as the pseudopotential \cite{RN3625}. A previous DFT study \cite{RN3824} systematically evaluated several van der Waals (vdW) correction schemes for describing interlayer interactions in InI and reported that the optB88-vdW functional \cite{RN3797, RN3798, RN4506} most accurately reproduced the experimental lattice parameters. Accordingly, we applied the optB88-vdW correction in all our calculations. The plane-wave cutoff energy was set to \SI{100}{\rydberg}, and the Monkhorst--Pack $k$-point meshes of $10 \times 10 \times 15$ and $7 \times 12 \times 6$ were used for \ce{InI} and low-pressure \ce{InI3}, respectively. The optimized lattice parameters and internal atomic coordinates (Table \ref{tab1}) showed good agreement with previously reported values \cite{RN4490, RN3645, RN3824, RN3780, RN3593, RN4141, RN4142}.

\begin{table}[h]
	\caption{Lattice parameters and internal atomic coordinates of the optimized \ce{InI} and low-pressure \ce{InI3} structures.}
	\label{tab1}
	\centering
	\begin{tabular}{ll}
	\hline
	Model & Lattice parameters and internal coordinates \\
	\hline 
   	\ce{InI} & $a = \SI{4.81243}{\angstrom}$ \\
    	$Cmcm$ & $b = \SI{12.94782}{\angstrom}$ \\
    	(No. 63, $Z = 4$)& $c = \SI{4.90259}{\angstrom}$\\
    	& \ce{In} = (0, \num{0.39609}, 3/4) \\
    	& \ce{I} = (0, \num{0.14356}, 3/4) \\
	\hline
    	\ce{InI3} & $a = \SI{9.89856}{\angstrom}$\\
    	$P2_1/c$ & $b = \SI{6.05283}{\angstrom}$ \\
    	(No. 14, $Z = 4$)&  $c = \SI{12.20190}{\angstrom}$ \\
    	& $\beta = \ang{107.5279}$ \\
    	& \ce{In} = (\num{0.20873}, \num{0.49811}, \num{0.55110}) \\ 
    	& \ce{I}(1) = (\num{0.99936}, \num{0.24450}, \num{0.87498}) \\
    	& \ce{I}(2) = (\num{0.33721}, \num{0.21661}, \num{0.72324}) \\ 
    	& \ce{I}(3) = (\num{0.34189}, \num{0.72413}, \num{0.94377}) \\
	\hline
	\end{tabular}
\end{table}

To compute the harmonic interatomic force constants (IFCs), we employed the finite-displacement method using the \texttt{Phonopy} package \cite{RN3234, RN3232}, applied to $4 \times 4 \times 6$ and $2 \times 4 \times 2$ supercells based on the primitive unit cells of \ce{InI} and low-pressure \ce{InI3}, respectively. Long-range dipole–dipole interactions were accounted for by incorporating the nonanalytic term into the dynamical matrices \cite{RN713}. The dielectric tensors and Born effective charges required for this correction were obtained through density functional perturbation theory calculations. Phonon transport in indium iodides was analyzed by solving the Peierls--Boltzmann (PB) transport equation under the single-mode relaxation time approximation (RTA) \cite{RN1908, RN1532}. The thermal conductivity tensor ($\kappa_{\text{PB}}$) for the Cartesian components $\alpha$ and $\beta$ is given by
\begin{equation}
	\displaystyle \kappa_\text{PB}^{\alpha\beta} = \sum_{\mu}c_{\mu}v_{\mu}^{\alpha}v_{\mu}^{\beta}\tau_{\mu}, \label{eq1}
\end{equation}
where $c_{\mu}$, $v_{\mu}$, and $\tau_{\mu}$ denote the volumetric specific heat, group velocity, and relaxation time of phonon $\mu$, respectively. Because of the weak \ce{In}--\ce{I} bonding and large atomic masses, these materials exhibit multiple low-frequency flat-band modes, which can enhance higher-order phonon scattering processes. Accordingly, we included both three-phonon and four-phonon scatterings in our calculations. Denoting $\tau_{\text{3ph}}$, $\tau_{\text{4ph}}$, and $\tau_{\text{iso}}$ as the relaxation times due to three-phonon, four-phonon, and phonon--isotope scatterings, respectively, the total relaxation time was evaluated using the spectral Matthiessen’s rule: $\tau_{\mu}^{-1} = \tau_{\text{3ph}, \mu}^{-1} + \tau_{\text{4ph}, \mu}^{-1} + \tau_{\text{iso}, \mu}^{-1}$. Phonon--isotope scattering was described using the Tamura model \cite{RN1557} with natural isotope concentration. Because $\tau_{\text{4ph}}$ calculation \cite{RN3630} was computationally demanding, particularly for systems with large unit cells and dense $q$-point meshes, we adopted an efficient sampling-based approach combined with maximum likelihood estimation, as proposed by Guo et al. \cite{RN3801}, to reduce computational costs.

To calculate anharmonic IFCs, we used \texttt{thirdorder.py} \cite{RN2032} and \texttt{Fourthorder.py} \cite{RN3823}, applying them to $2 \times 2 \times 3$ and $1 \times 2 \times 1$ supercells for \ce{InI} and \ce{InI3}, respectively. Considering computational costs, we limited the fourth-order anharmonic IFCs to interactions within nearest-neighbor (NN) atoms for both materials. In contrast, for the third-order anharmonic IFCs, we set cutoff radii of 5.5 and \SI{4.3}{\angstrom} for \ce{InI} and \ce{InI3}, respectively, to capture interactions between adjacent \ce{NaCl}-type double layers in InI and between \ce{In2I6} dimers in \ce{InI3}. Phonon transport properties were computed using the \texttt{ShengBTE} package and its extensions \cite{RN2032, RN3823}. Convergence tests on $q$-point meshes confirmed that $16 \times 16 \times 16$ and $8 \times 16 \times 8$ $q$-point meshes for \ce{InI} and low-pressure \ce{InI3}, respectively, yield well-converged thermal conductivities. Furthermore, a comparison between the full solution and RTA for the linearized Boltzmann transport equation showed a negligible difference (less than 1\%). Therefore, RTA was used for all subsequent transport calculations. To efficiently evaluate $\tau_{\text{4ph}}$, we adopted sampling numbers that reliably reproduced the thermal conductivities obtained from the full Brillouin zone on coarse meshes and applied these sampling parameters to the denser mesh calculations. The effects of the third-order IFC cutoff radii and $q$-mesh size on thermal conductivity are detailed in Fig. \ref{figS1} and Table \ref{tabS1} in the Supplementary Materials.

\section{\label{sec:results_and_discussion}Results and Discussion}
\subsection{\label{subsec:InI_and_InI3}Phonon transport characteristics of \ce{InI} and low-pressure \ce{InI3}}

\begin{figure*}[t!]
	\centering
	\includegraphics[width=0.95\textwidth]{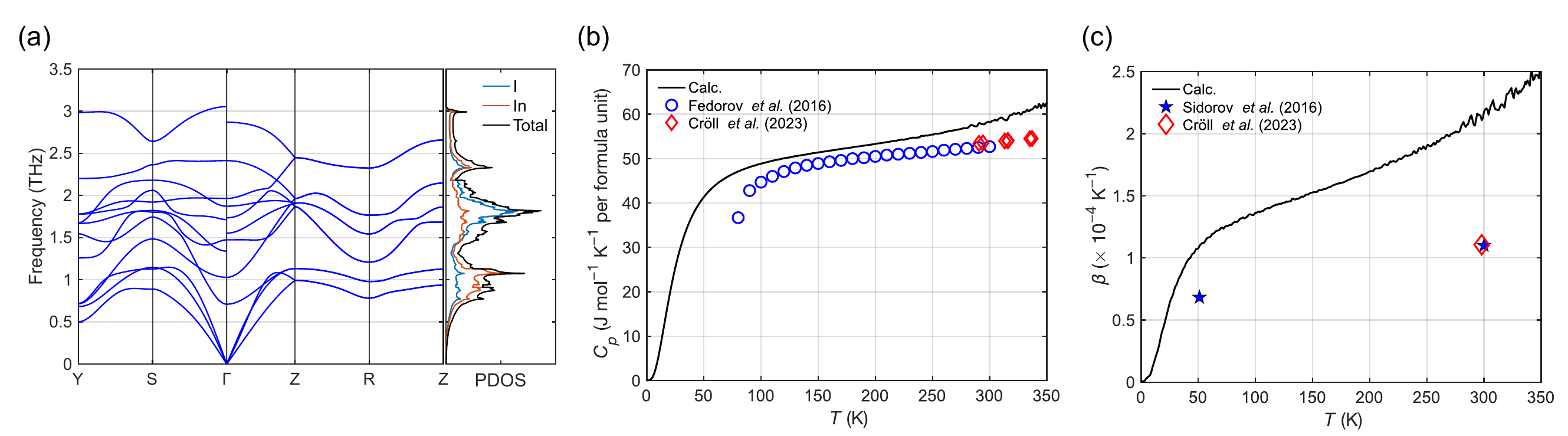}
	\caption{
	Vibrational and thermodynamic properties of \ce{InI}. (a) Phonon dispersion relation along high-symmetry lines in the Brillouin zone and partial density of states. (b) Temperature dependence of the constant-pressure specific heat per formula unit. (c) Temperature dependence of the volumetric thermal expansion coefficient. The markers in panels (b) and (c) denote experimental data \cite{RN4032, RN4033,  RN3780}.
	}
	\label{fig2}
\end{figure*}

Figure \ref{fig2}(a) presents the calculated phonon dispersion relation and partial density of states (PDOS) of \ce{InI}. Because of the large atomic masses and weak \ce{In}--\ce{I} bonding, the overall phonon frequencies were suppressed. In particular, similar to vdW crystals \cite{RN1067}, the acoustic phonons were confined below \SI{1}{\tera \hertz}, indicating intrinsically low thermal conductivity in \ce{InI}. Beyond thermal conductivity, other thermophysical properties of \ce{InI}, such as heat capacity and thermal expansion, have also been extensively investigated. To validate our calculations, we evaluated the constant-pressure specific heat per formula unit and the volumetric thermal expansion coefficient using the quasi-harmonic approximation \cite{RN4507}. As shown in Fig. \ref{fig2}(b), the calculated temperature dependence of specific heat exhibited reasonable agreement with the experimental data over a wide temperature range \cite{RN4032, RN3780}, although some discrepancies were observed below \SI{100}{\kelvin} and above \SI{300}{\kelvin}. In contrast, the calculated volumetric thermal expansion coefficients were approximately twice as high as the reported experimental values \cite{RN4033,  RN3780}---reaching \SI{2e-4}{\per \kelvin} at \SI{300}{\kelvin}---comparable to that of molten \ce{InI} (Fig. \ref{fig2}(c)). The relatively large fluctuations in the high-temperature range for both the specific heat and the thermal expansion coefficient might have arisen from the limited number of volumes used in the Gibbs free energy calculations. Additionally, although previous studies have reported anisotropic, temperature-dependent linear thermal expansion coefficients along different crystal axes, our calculations maintained fixed ratios between the lattice parameters. Given the strong lattice anharmonicity of the \ce{InI} crystal, the incorporation of anisotropic structural effects into lattice anharmonicity evaluation will be crucial for future work \cite{RN3077, RN4508}.

\begin{figure}[t!]
	\centering
	\includegraphics[width=0.4\textwidth]{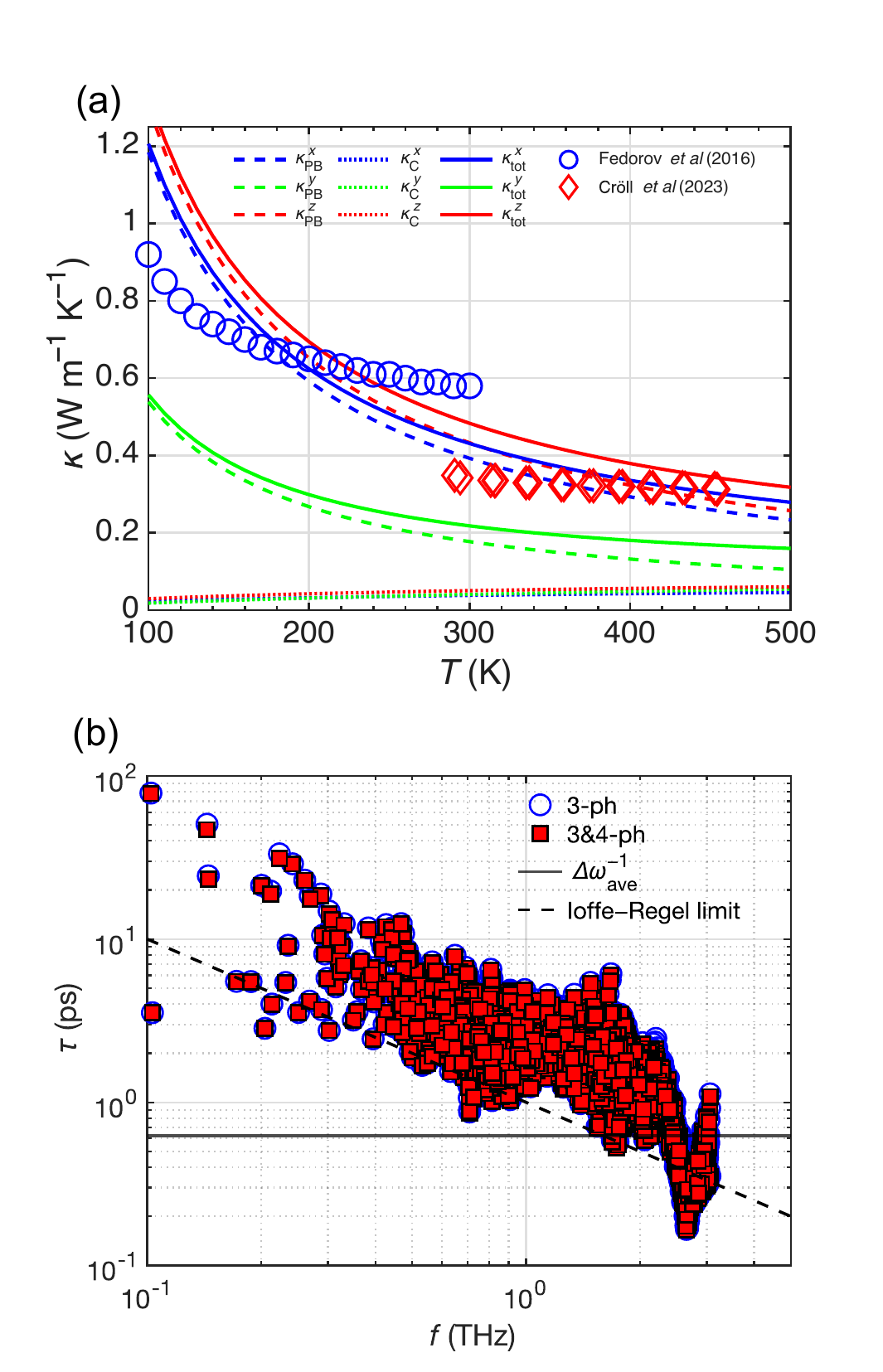}
	\caption{
	Phonon transport characteristics of \ce{InI}. (a) Temperature dependence of thermal conductivity ($\kappa$) along the Cartesian directions. The dashed and dotted lines represent the particle-like Peierls (diagonal) and wave-like interband tunneling (off-diagonal) contributions to $\kappa$, denoted as $\kappa_{\text{PB}}$ and $\kappa_{\text{C}}$, respectively. The solid line indicates the total thermal conductivity ($\kappa_{\text{tot}} = \kappa_{\text{PB}} + \kappa_{\text{C}}$). The blue open circles and red open diamonds indicate the experimental results \cite{RN4032, RN3780}. (b) Frequency-dependent relaxation times ($\tau$) at \SI{300}{\kelvin}. The blue open circles and red filled squares represent $\tau$ including only three-phonon and three- and four-phonon scatterings, respectively. The solid and dashed lines correspond to the inverse of the average interband spacing ($\varDelta \omega_{\text{ave}}^{-1}$) \cite{RN2728} and the Ioffe--Regel limit ($2\pi / \omega$) \cite{RN1482}, respectively. Phonon--isotope scattering \cite{RN1557} was included in all calculations.
	}
	\label{fig3}
\end{figure}

Figure \ref{fig3}(a) illustrates the temperature dependence of the calculated thermal conductivities of InI along the Cartesian directions. At \SI{300}{\kelvin}, the thermal conductivity along the \ce{NaCl}-type double-layer stacking direction (i.e., $y$-direction) was \SI{0.18}{\watt \per \meter \per \kelvin}, whereas that along the direction perpendicular to the stacking was \SI{0.4}{\watt \per \meter \per \kelvin}. This anisotropic behavior reflects the inherent structural anisotropy of \ce{InI}. The in-plane thermal conductivity within the \ce{NaCl}-type double layer was higher than the previously reported value for monolayer \ce{InI} (\SI{0.27}{\watt \per \meter \per \kelvin}) \cite{RN3787}. For comparison, Fig. \ref{fig3}(a) also shows experimental thermal conductivities measured along the $b$-axis (stacking direction) using the longitudinal heat flux and Xenon laser flush methods \cite{RN4032, RN3780}. The discrepancies between these measurements could be attributed to the differences in apparent thermal conductivity arising from the infrared transparency of \ce{InI} and the specific characteristics of each measurement technique. Whereas the experimental values showed a gradual decrease with increasing temperature, the calculated thermal conductivities exhibited a steeper temperature dependence. Given the low thermal conductivity of InI, wave-like interband tunneling may have contributed non-negligibly to the overall thermal transport \cite{RN1709, RN2728}. To assess this, we evaluated the wave-like contribution using the following expression \cite{RN2728}:
\begin{align}
	\displaystyle \kappa_{\text{C}}^{\alpha\beta} &= \sum_{\mu,\mu^\prime(\not=\mu)}\frac{\omega_{\mu}+\omega_{\mu^\prime}}{4}\left(\frac{c_{\mu}}{\omega_{\mu}}+\frac{c_{\mu^\prime}}{\omega_{\mu^\prime}}\right)v_{\mu\mu^\prime}^{\alpha}v_{\mu\mu^\prime}^{\beta} \notag \\
	& \times \frac{[\varGamma_{\mu}+\varGamma_{\mu^\prime}]/2}{[\omega_{\mu}-\omega_{\mu^\prime}]^2+[\varGamma_{\mu}+\varGamma_{\mu^\prime}]^2/4}, \label{eq2}
\end{align}
where $v_{\mu\mu^\prime}$ is the generalized group velocity between two phonons and $\varGamma_{\mu}$ is the total scattering rate. The calculated wave-like contribution ($\kappa_{\text{C}}$) was $\sim 10$\% of the particle-like Perierls contribution ($\kappa_{\text{PB}}$), which was reasonable because the $\tau$ of phonons exceeded the inverse of the average interband spacing ($\varDelta \omega_{\text{ave}}^{-1}$) \cite{RN2728}, as shown in Fig. \ref{fig3}(b) (see also Fig. \ref{figS2}). Furthermore, $\kappa_{\text{C}}$ contributed minimally to the temperature dependence of the total thermal conductivity and therefore could not account for the discrepancy in temperature trends between our calculations and the experimental data. Additionally, three-phonon scattering was found to dominate phonon transport, and the $\tau$ of most phonons exceeded the Ioffe--Regel limit \cite{RN1482}, supporting the validity of the quasi-particle picture of phonons in \ce{InI}. The notable disagreement in the temperature dependence might have stemmed instead from the treatment of lattice anharmonicity and its coupling with structural anisotropy.

\begin{figure*}[t!]
	\centering
	\includegraphics[width=0.95\textwidth]{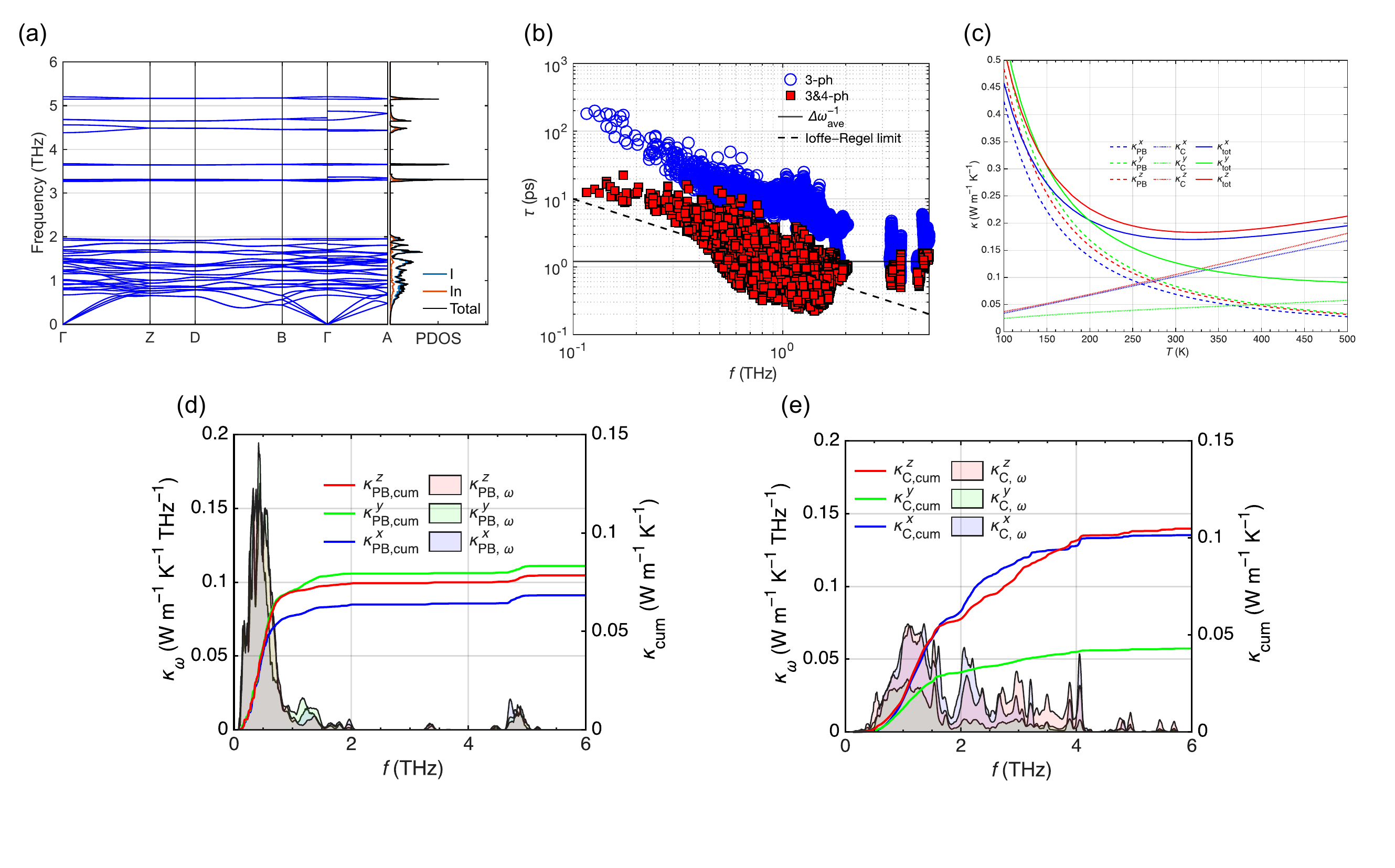}
	\caption{
	Phonon transport characteristics of low-pressure \ce{InI3}. (a) Phonon dispersion relation and PDOS. (b) Frequency-dependent $\tau$ at \SI{300}{\kelvin}. The definitions of the markers and lines are the same as those in Fig. \ref{fig3}. (c) Temperature dependence of $\kappa_{\text{PB}}$,  $\kappa_{\text{C}}$, and $\kappa_{\text{tot}}$ along the Cartesian directions. (d, e) Spectral thermal conductivities ($\kappa_{\omega}$, left axis) and their accumulations ($\kappa_{\text{cum}}$, right axis) at  \SI{300}{\kelvin} along the Cartesian directions for the contributions of $\kappa_{\text{PB}}$ and $\kappa_{\text{C}}$, respectively.
	}
	\label{fig4}
\end{figure*}

We next examined the phonon transport properties of low-pressure \ce{InI3}. Figure \ref{fig4}(a) shows the phonon dispersion relation and PDOS of this material. Although the maximum phonon frequency of \ce{InI3} was higher than that of \ce{InI}, the presence of multiple flat-band optical modes suggests an even lower thermal conductivity. As shown by the frequency-dependent $\tau$ at \SI{300}{\kelvin} (Fig. \ref{fig4}(b)), the inclusion of four-phonon scattering reduced $\tau$ by nearly an order of magnitude compared with calculations that considered only three-phonon scatterings. Moreover, many phonons above \SI{0.5}{\tera \hertz} exhibited $\tau$ shorter than $\varDelta \omega_{\text{ave}}^{-1}$, indicating the potential importance of wave-like contributions.

Figure \ref{fig4}(c) displays the calculated temperature-dependent thermal conductivity of low-pressure \ce{InI3}. Despite its structural anisotropy, $\kappa_{\text{PB}}$ exhibited relatively weak anisotropy. This result might have been due to four-phonon scatterings, which preferentially suppress heat-carrying phonons with directionally dependent group velocities. In fact, $\kappa_{\text{PB}}$ calculated using only three-phonon scatterings exhibited notable anisotropy (Fig. \ref{figS1}). Additionally, whereas $\kappa_{\text{PB}}$ with only three-phonon scattering was $\sim$\SI{0.5}{\watt \per \meter \per \kelvin} at \SI{300}{\kelvin}, the inclusion of four-phonon scattering reduced this value to less than one-fifth. We also computed the temperature dependence of $\kappa_{\text{C}}$ (Fig. \ref{fig4}(c)), which interestingly became comparable to $\kappa_{\text{PB}}$ at around \SI{270}{\kelvin} and dominated at higher temperatures.

Figures \ref{fig4}(d) and \ref{fig4}(e) present the spectral thermal conductivities for $\kappa_{\text{PB}}$ and $\kappa_{\text{C}}$ at \SI{300}{\kelvin}, respectively. For $\kappa_{\text{PB}}$, low-frequency phonons below \SI{1}{\tera \hertz} were the dominant contributors, whereas $\kappa_{\text{C}}$ received substantial contributions from phonons across a broader frequency range. In the low-frequency region, however, $\kappa_{\text{PB}}$ remained the dominant term---consistent with the fact that $\tau$ values in such a frequency region are longer than $\varDelta \omega_{\text{ave}}^{-1}$. Because low-pressure \ce{InI3} can be considered a molecular crystal composed of \ce{In2I6} dimers \cite{RN3645, RN4487}, its intrinsically low thermal conductivity on the order of \SI{0.1}{\watt \per \meter \per \kelvin} is intuitively reasonable, comparable to that of amorphous and polymeric materials. However, the presence of many phonons with $\tau$ values shorter than the Ioffe--Regel limit suggests that the quasi-particle picture of phonons may not be strictly valid. Accordingly, further experimental investigations, particularly the measurement of temperature-dependent thermal conductivity, are essential to validate the present theoretical evaluations.

\subsection{\label{subsec:high-pressure_InI3}Structures of high-pressure \ce{InI3} and their phonon transports}
\begin{figure*}[t!]
	\centering
	\includegraphics[width=0.95\textwidth]{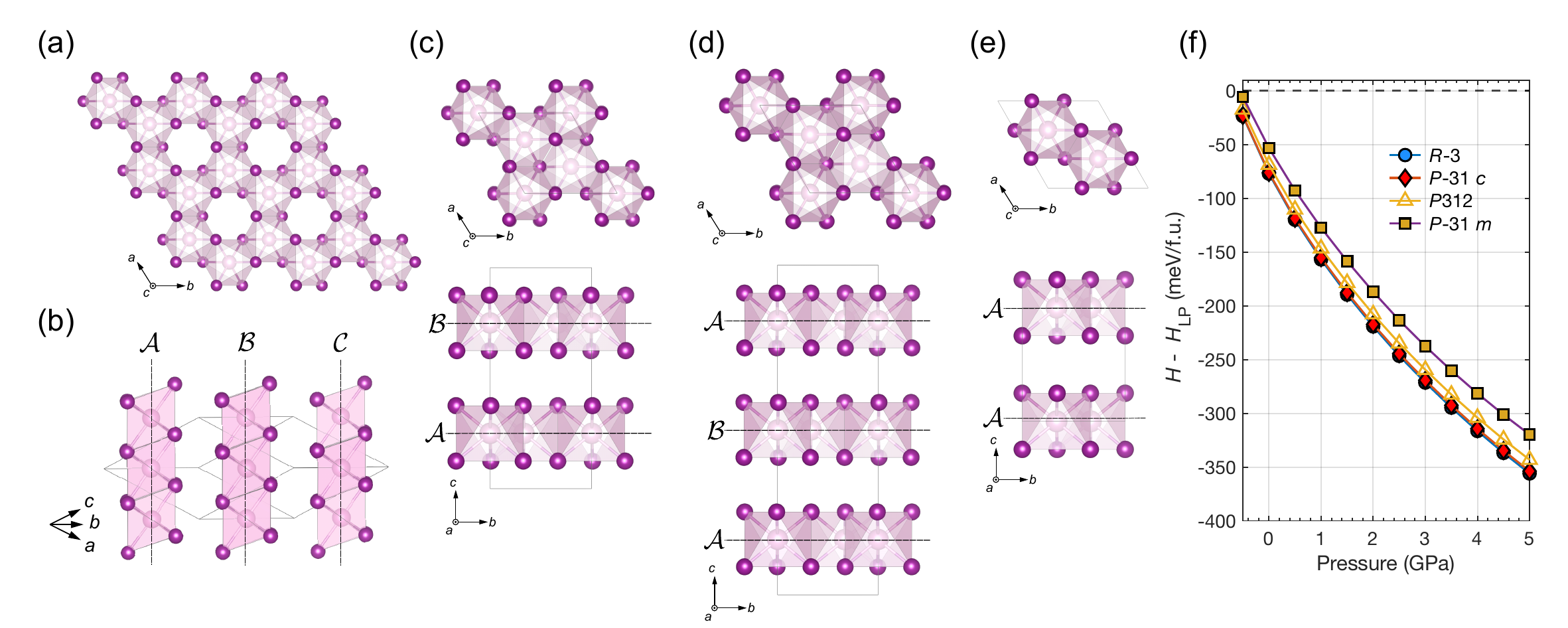}
	\caption{
	 Schematics of the (a) edge-sharing \ce{In2I6} monolayer, (b) $R\bar{3}$, (c) $P\bar{3}1c$, (d) $P312$, and (e) $P\bar{3}1m$  high-pressure \ce{InI3} models. $\mathcal{A}$, $\mathcal{B}$, and $\mathcal{C}$ indicate layer indexes. The structures are visualized using \texttt{VESTA} \cite{RN3640}. (f) Pressure dependence of enthalpies per formula unit for high-pressure phases relative to that of the low-pressure phase.
	}
	\label{fig5}
\end{figure*}

The previously reported rhombohedral \ce{InI3} structure formed under pressure treatment exhibits disordered indium sites, with occupancies of 26.7\% and 73.3\% at the $3a$ and $6c$ Wyckoff positions, respectively \cite{RN3593}. To explore possible ordered configurations, we constructed several structural models with fully ordered indium sites. To preserve the \ce{InI3} stoichiometry, three of the nine total indium sites (i.e., the $3a$ and $6c$ sites combined) must be removed. When all three $3a$ sites corresponding to the fractional coordinates $(0,0)$, $(1/3,2/3)$ and $(2/3,1/3)$ on the $ab$-plane were removed, we obtained a structure with an $R\bar{3}$ space group (No. 148, $Z = 2$), where three edge-sharing \ce{In2I6} layers with different vacancy positions (Fig. \ref{fig5}(a)) were $\mathcal{ABC}$-stacked along the $c$-axis. In contrast, removing the $(0,0)$ $3a$ site and the two vertically aligned $6c$ sites above it shortened the $c$-axis lattice constant and yielded a structure with $P\bar{3}1m$ symmetry (No. 162, $Z = 2$), where \ce{In2I6} layers formed $\mathcal{AA}$ stacking. Using the same construction strategy, we also generated $\mathcal{AAB}$-stacked $P312$ (No. 149, $Z = 2$) and $\mathcal{AB}$-stacked $P\bar{3}1c$ (No. 163, $Z = 4$) structural models. The optimized lattice parameters and internal coordinates for these four high-pressure models, along with schematic illustrations, are summarized in Table \ref{tab2} and Fig. \ref{fig5}(b--e).

\begin{table}[h]
	\caption{Lattice parameters and internal atomic coordinates of the optimized high-pressure \ce{InI3} models at \SI{0.1}{\mega \pascal}.}
	\label{tab2}
	\centering
	\begin{tabular}{ll}
	\hline
	Model & Lattice parameters and internal coordinates \\
	\hline 
   	$R\bar{3}$ & $a = \SI{8.06180}{\angstrom}$ \\
	(No. 148, $Z=2$) & $\alpha = \SI{53.9726}{\degree}$ \\
    	& \ce{In} = (\num{0.16609}, \num{0.16609}, \num{0.16609}) \\
    	& \ce{I} = (\num{0.58168}, \num{0.92513}, \num{0.24414}) \\
	\hline
   	$P\bar{3}1c$ & $a = \SI{7.31533}{\angstrom}$ \\
	(No. 163, $Z=4$) & $c = \SI{13.74844}{\angstrom}$\\
	& $\gamma = \SI{120}{\degree}$ \\
    	& \ce{In}(1) = (0, 0, 1/4), In(2)=(2/3, 1/3, 1/4) \\
    	& \ce{I} = (\num{0.33386}, \num{0.00739}, \num{0.37534}) \\
	\hline
   	$P312$ & $a = \SI{7.30923}{\angstrom}$ \\
	(No. 149, $Z=2$) & $c = \SI{20.70872}{\angstrom}$\\
	& $\gamma = \SI{120}{\degree}$ \\
    	& \ce{In}(1) = (0, 0, \num{0.83193}), In(2)=(2/3, 1/3, 1/2) \\
    	& \ce{In}(3) = (1/3, 2/3, \num{0.83250}), In(4)=(0, 0, 1/2) \\	
    	& \ce{I}(1) = (\num{0.66739}, \num{0.67171}, \num{0.91500}) \\
    	& \ce{I}(2) = (\num{0.32612}, \num{0.99154}, \num{0.25123}) \\
    	& \ce{I}(3) = (\num{0.99275}, \num{0.32663}, \num{0.58333}) \\
	\hline
   	$P\bar{3}1m$ & $a = \SI{7.29639}{\angstrom}$ \\
	(No. 162, $Z=2$) & $c = \SI{6.96631}{\angstrom}$\\
	& $\gamma = \SI{120}{\degree}$ \\
    	& \ce{In} = (1/3, 2/3, 0) \\
    	& \ce{I} = (0, \num{0.66141}, \num{0.75303}) \\
	\hline	
	\end{tabular}
\end{table}

Figure \ref{fig5}(f) shows the computed pressure dependence of the enthalpy differences between the high-pressure models and the low-pressure \ce{InI3}. Because the exact pressure conditions for the phase transformation were not specified in the previous experimental study---which reported the appearance of the high-pressure phase upon mechanical gliding---we evaluated the enthalpy over a pressure range of $-0.5$--\SI{5}{\giga \pascal}. Within this range, all four high-pressure structures exhibited lower enthalpy than the low-pressure phase, indicating that these structures are energetically favorable even at ambient pressure. However, pressure values estimated from DFT calculations can vary depending on the choice of the exchange--correlation functional, especially near ambient conditions. Among the high-pressure models, $R\bar{3}$ was the most stable, followed by $P\bar{3}1c$, $P312$, and $P\bar{3}1m$. This trend suggests that stacking \ce{In2I6} layers with different vacancy positions enhances energetic stability. Nevertheless, the enthalpy differences among the high-pressure models were within \SI{30}{\milli \electronvolt} per formula unit, implying that multiple stacking configurations may coexist under finite-temperature conditions.

\begin{figure*}[]
	\centering
	\includegraphics[width=0.95\textwidth]{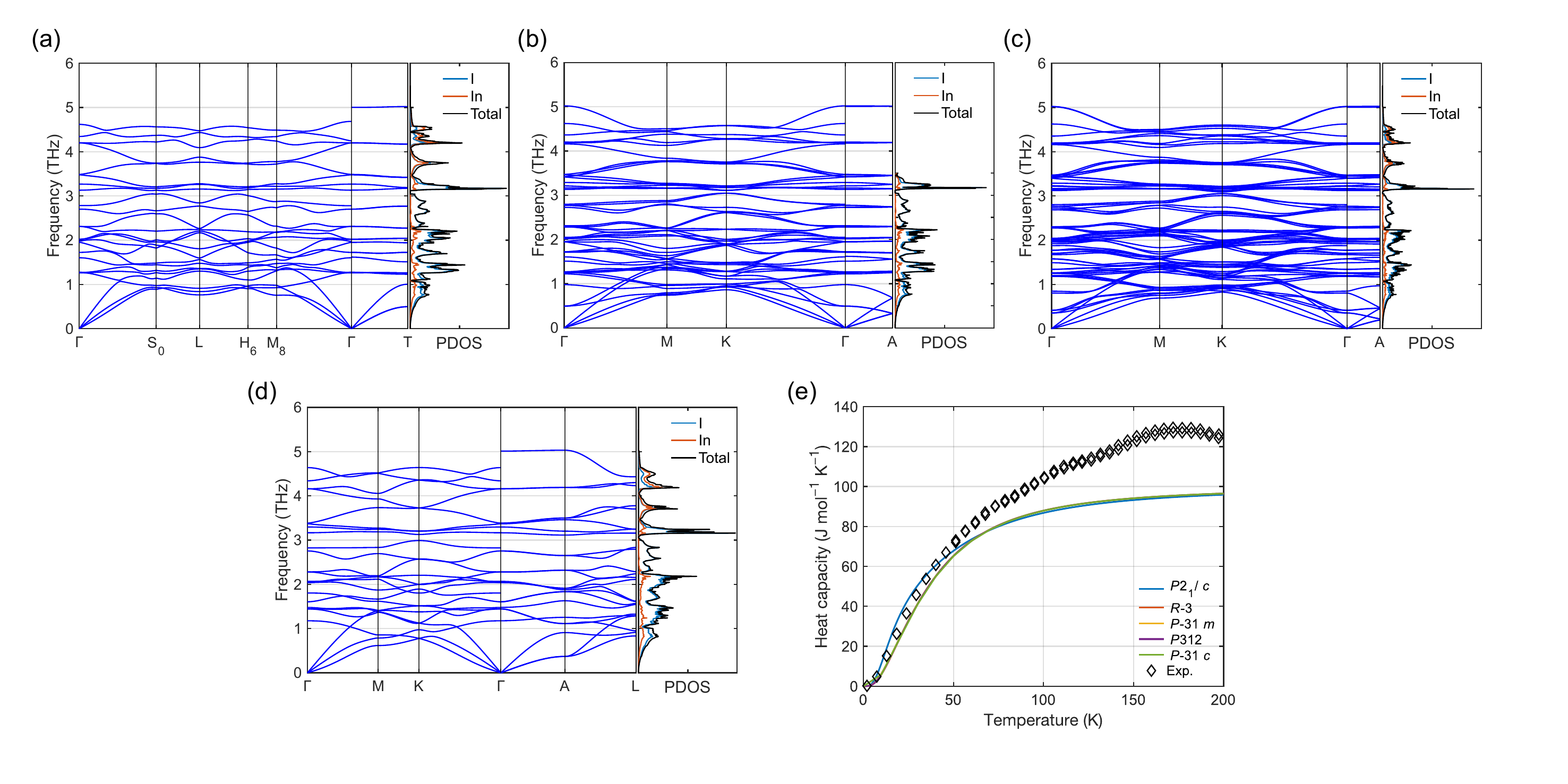}
	\caption{
	Phonon dispersion relations and PDOSs of high-pressure \ce{InI3} with the crystalline phases of (a) $R\bar{3}$, (b) $P\bar{3}1c$, (c) $P312$, and (d) $P\bar{3}1m$. (e) Temperature-dependent volumetric specific heats per formula unit for high-pressure and low-pressure phases. The black open diamond denotes the experimental measurement \cite{RN3593}.
	}
	\label{fig6}
\end{figure*}

Although all high-pressure models exhibited comparable energetic stability, their structural differences may influence thermophysical properties such as heat capacity and thermal conductivity. Figure \ref{fig6}(a--d) shows the phonon dispersion relations and PDOSs for the four models, obtained from first-principles calculations. For the calculation of harmonic IFCs, we constructed supercells based on the optimized lattice parameters listed in Table \ref{tab2}. Specifically, we used $3 \times 3 \times 3$ supercells for the $R\bar{3}$ and $P\bar{3}1m$ models, a $3 \times 3 \times 1$ supercell for $P312$, and a $3 \times 3 \times 2$ supercell for $P\bar{3}1c$. Although the number of phonon modes differed among the models because of the differences in the number of atoms per unit cell, the overall features of their phonon dispersion relations were qualitatively similar.

Figure \ref{fig6}(e) presents the temperature dependence of the volumetric specific heat per formula unit for each model. The magnitude and temperature dependence of the specific heat were nearly identical across all models. This similarity could be attributed to their similar vibrational characteristics, as indicated by the PDOS results (Fig. \ref{fig6}(a--d)). When compared with the low-pressure \ce{InI3}, the high-pressure phases exhibited a slower increase in specific heat at low temperatures but converged to nearly the same values at higher temperatures. Furthermore, comparison with the previously reported experimental data indicated that the measured values fell between the calculated results for the high- and low-pressure phases, although noticeable discrepancies appeared above \SI{50}{\kelvin}.

A careful comparison of the phonon dispersion relations revealed that group velocities and the features of flat-band modes depended on the structural model. However, as suggested by the similarity in specific heat, the thermal conductivity differences among the high-pressure models were expected to be minor. As with the low-pressure \ce{InI3}, we set the cutoff radii of third-order IFCs to 5--\SI{6}{\angstrom} while limiting fourth-order IFC interactions to NN atoms for the high-pressure models (see Table \ref{tabS1} and Figs. \ref{figS3} and \ref{figS4}).

\begin{figure}[t!]
	\centering
	\includegraphics[width=0.4\textwidth]{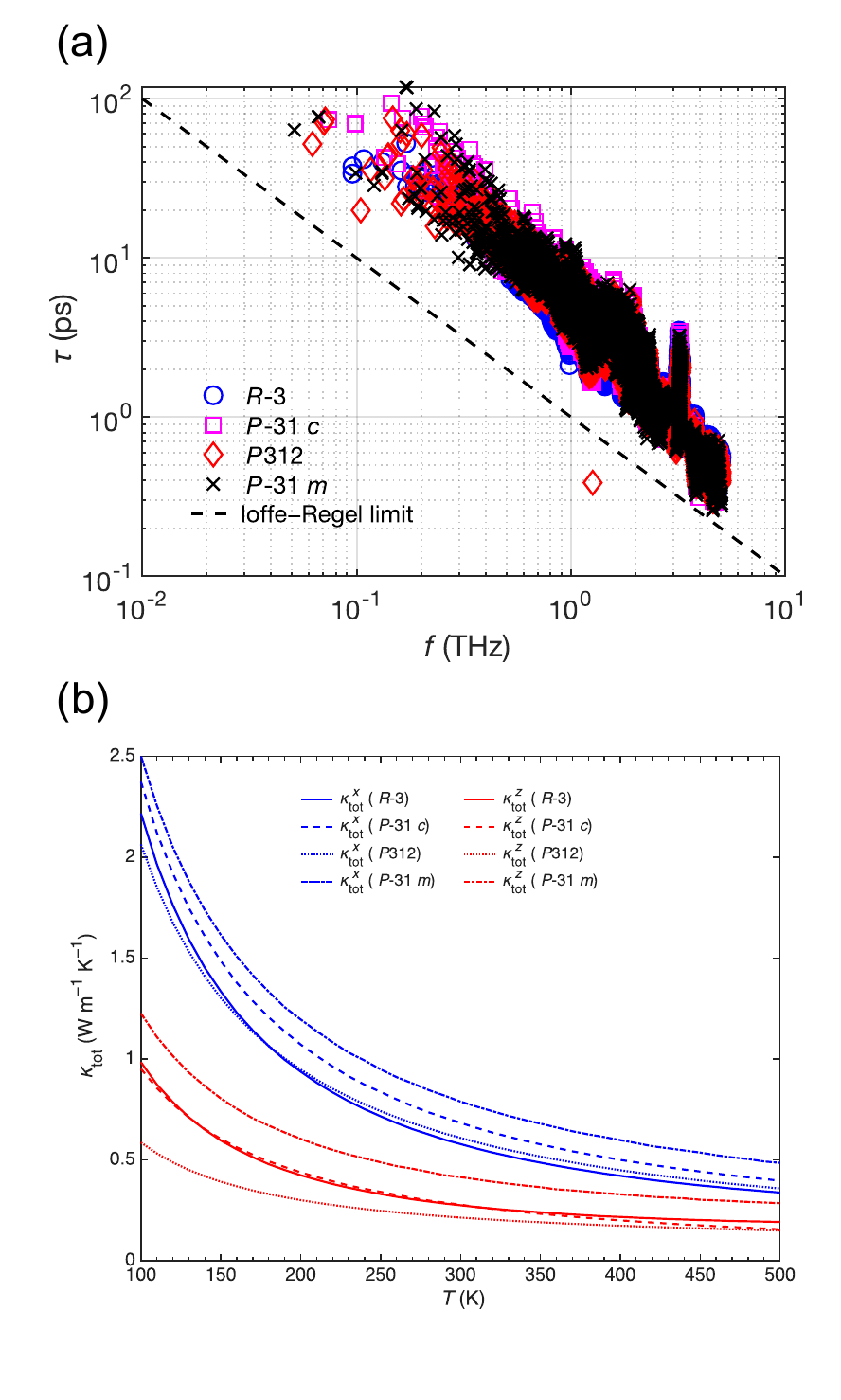}
	\caption{Phonon transport characteristics of high-pressure \ce{InI3} (a) Frequency-dependent $\tau$ at \SI{300}{\kelvin}. The dashed line indicates the Ioffe--Regel limit ($2\pi/\omega$). (b) Temperature-dependent $\kappa_{\text{tot}}$ along the Cartesian directions.}
	\label{fig7}
\end{figure}

Figure \ref{fig7}(a) compares the frequency-dependent $\tau$ at \SI{300}{\kelvin} for the different models (Fig. \ref{figS5}). Although slight differences in $\tau$ were observed, the frequency dependence and order of magnitude were similar across all structures, indicating that phonon scattering characteristics are nearly independent of the structural model. Furthermore, because the $\tau$ values of many phonons exceeded $\varDelta \omega_{\text{ave}}^{-1}$---which varied slightly among models because of the differences in the number of atoms per unit cell but was generally around \SI{2}{\pico \second}---particle-like contributions ($\kappa_{\text{PB}}$) were expected to dominate thermal transport. In fact, $\kappa_{\text{C}}$ at 300 K was less than 20\% of $\kappa_{\text{PB}}$ for all high-pressure models. Although the $\kappa_{\text{C}}/\kappa_{\text{PB}}$ ratio varied slightly depending on the structural model and transport direction, its overall contribution was minor relative to the low-pressure \ce{InI3} (Fig. \ref{figS6}).

Figure \ref{fig7}(b) illustrates the temperature dependence of $\kappa_{\text{tot}}$, including the $\kappa_{\text{C}}$ contributions, enabling direct comparison with the low-pressure phase. In-plane heat conduction within the \ce{In2I6} layers was isotropic across all high-pressure models. Although the differences between in-plane and out-of-plane $\kappa_{\text{tot}}$ for the $R\bar{3}$ and $P\bar{3}1m$ models reached about 35\% and 50\% at \SI{300}{\kelvin}, respectively, the overall magnitude, anisotropy, and temperature dependence of $\kappa_{\text{tot}}$ were qualitatively similar among the four high-pressure structures. This conclusion was further supported by the spectral thermal conductivity results shown in Fig. \ref{figS7}.

\begin{figure*}[]
	\centering
	\includegraphics[width=0.95\textwidth]{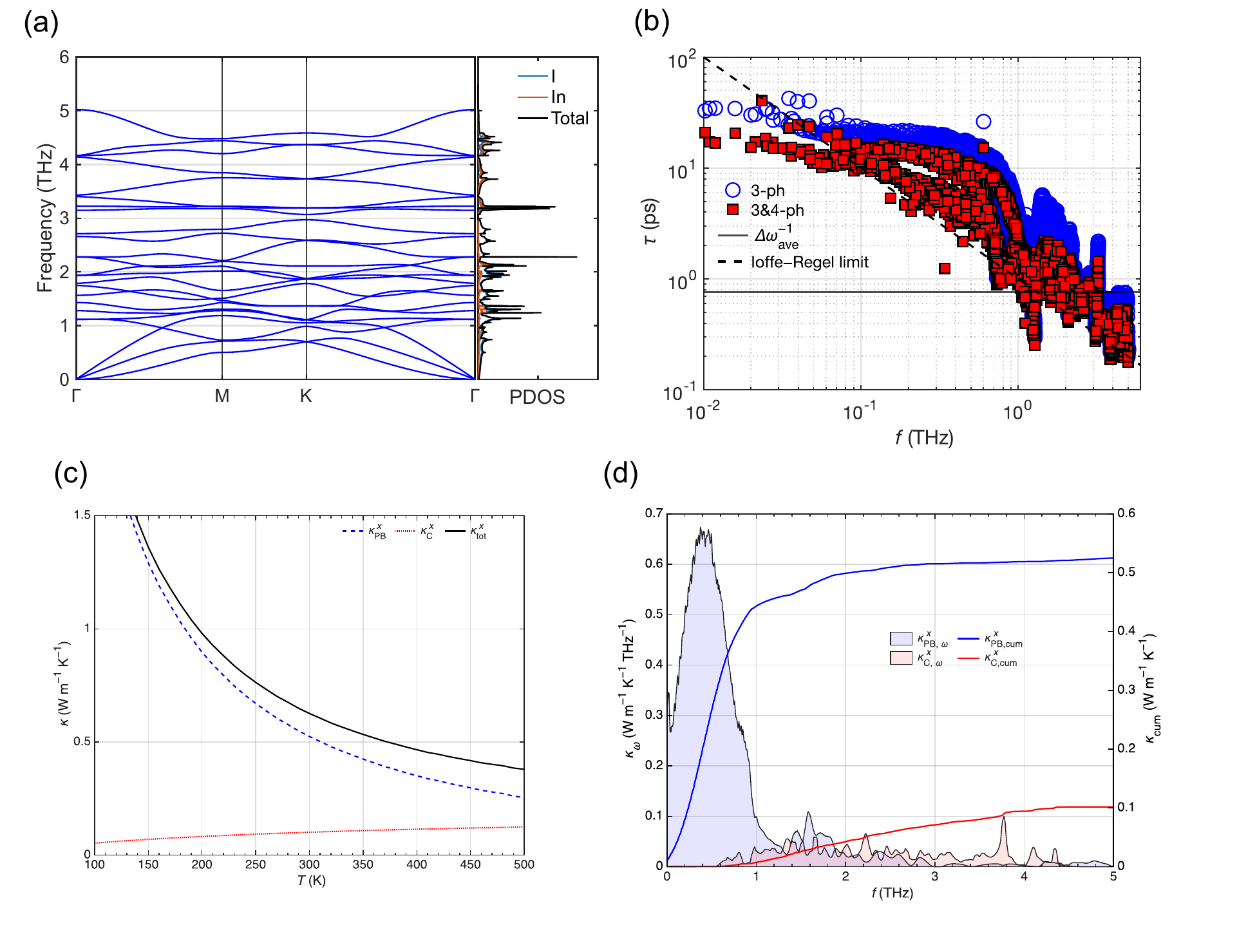}
	\caption{Phonon transport characteristics of monolayer \ce{In2I6}: (a) Phonon dispersion relation and PDOS. (b) Frequency-dependent $\tau$ at \SI{300}{\kelvin}. The definitions of the markers and lines are the same as those in Fig. \ref{fig3}. (c) Temperature dependence of $\kappa_{\text{tot}}$, $\kappa_{\text{PB}}$, and $\kappa_{\text{C}}$, along the Cartesian directions. (d) Spectral thermal conductivities ($\kappa_{\omega}$, left axis) and their accumulations ($\kappa_{\text{cum}}$, right axis) at \SI{300}{\kelvin} along the Cartesian directions.}
	\label{fig8}
\end{figure*}

Because of the weak interactions between \ce{In2I6} layers, intrinsic phonon transport within each layer likely governed the in-plane thermal conductivity in the high-pressure models. To verify this, we investigated phonon transport in a monolayer \ce{In2I6}. This structure was based on the $P\bar{3}1m$ symmetry, with a vacuum layer of \SI{10}{\angstrom} inserted along the out-of-plane direction to eliminate interactions due to periodic boundary conditions. The phonon dispersion relation and PDOS of the monolayer, calculated using a $4 \times 4$ supercell of the primitive unit cell, are shown in Fig. \ref{fig8}(a). Aside from hybridized phonon models, which likely originated from interlayer interactions in bulk structures, the overall dispersion closely resembled the in-plane vibrational characteristics of the four high-pressure models.

Figure \ref{fig8}(b) shows the frequency-dependent $\tau$ value at \SI{300}{\kelvin} for the monolayer. As in the high-pressure phases, a cutoff radius of \SI{5.9}{\angstrom} and NN interactions were applied for third- and fourth-order IFCs, respectively (see Table \ref{tabS1} and Fig. \ref{figS8} for details). Despite the absence of interlayer interactions, the $\tau$ values were generally small, with four-phonon scattering having a substantial impact. Although many phonons exhibited $\tau$ values close to the Ioffe--Regel limit, we assumed the validity of the quasi-particle picture in evaluating the thermal conductivity. The monolayer thickness was set to \SI{6.94}{\angstrom}, corresponding to the average interlayer spacing in the high-pressure structures. The computed $\kappa_{\text{tot}}$, including both particle- and wave-like contributions, was \SI{0.6}{\watt \per \meter \per \kelvin} at \SI{300}{\kelvin}. The magnitude and the spectral distributions of the thermal conductivity (Fig. \ref{fig8}(d)) indicated that phonon transport in the monolayer was consistent with the in-plane behavior of the high-pressure phases.

In contrast, a recent machine-learning-based prediction considering only three-phonon scattering reported a much higher thermal conductivity of \SI{8.86}{\watt \per \meter \per \kelvin} at \SI{300}{\kelvin} for the monolayer, assuming a smaller thickness of \SI{3.35}{\angstrom} \cite{RN4468}. Even when the same thickness as that used in the machine learning model was adopted, our calculated $\kappa_{\text{tot}}$ remained approximately an order of magnitude smaller. Unlike materials such as graphene, where monolayering enhances out-of-plane symmetry \cite{RN1414}, the \ce{In2I6} monolayer inherently possesses structural complexity. This structural feature may not have been fully captured in previous machine learning explorations.

\section{\label{sec:conclusion}Conclusion}
In this study, we systematically investigated the phonon transport properties of layered \ce{InI} and molecular-crystal-like \ce{InI3} using first-principles anharmonic lattice dynamics calculations. Given their heavy constituent atoms and weak \ce{In}--\ce{I} bonding, both materials exhibited extremely low lattice thermal conductivities, well below \SI{1}{\watt \per \meter \per \kelvin} at \SI{300}{\kelvin}. For \ce{InI}, the phonon relaxation times exceeded the Ioffe--Regel limit, confirming the validity of the quasi-particle picture for phonons. Thermal transport was therefore predominantly governed by particle-like phonon transport described by the PB framework, whereas the wave-like interband tunneling contribution accounted for only about 10\% of the total thermal conductivity. In contrast, the low-pressure \ce{InI3} exhibited much stronger anharmonicity, where four-phonon scattering significantly reduced phonon lifetimes and suppressed the lattice thermal conductivity to less than one-fifth of the value obtained when only three-phonon scattering was considered. Moreover, above $\sim$\SI{270}{\kelvin}, the wave-like contribution to thermal transport became comparable to or larger than the particle-like contribution, indicating a qualitative change in the dominant heat conduction mechanism.

For the high-pressure phase of \ce{InI3}, several ordered structural models were constructed to account for the indium-site disorder reported experimentally in the rhombohedral phase. All proposed models were energetically more stable than the low-pressure phase, with enthalpy differences among them comparable to thermal energy, suggesting the possible coexistence of multiple stacking configurations. Phonon transport calculations revealed that the thermal conductivities of these high-pressure models were similar to each other, indicating that in-plane heat transport was primarily governed by the intrinsic vibrational properties of the \ce{In2I6} layers rather than the stacking sequence. Further, an isolated monolayer \ce{In2I6} exhibited an in-plane thermal conductivity comparable to that of the high-pressure phases. This behavior was distinct from that of simpler two-dimensional crystals, such as graphene, and was likely attributable to the intrinsic structural complexity and strong anharmonicity of \ce{In2I6}. A closer comparison between the monolayer and stacked systems showed that exfoliation did not significantly enhance the thermal conductivity and that differences remained in the frequency dependence of phonon relaxation times, particularly in the low-frequency regime. These observations suggest that, in such structurally complex layered materials, stacking may play a constructive role in thermal transport. Specifically, the confinement of individual layers by adjacent layers or the weak vdW interactions between neighboring layers may stabilize certain phonon modes or induce mode hybridization, thereby contributing positively to heat conduction. Overall, this study provides important insights into heat conduction mechanisms in complex layered and molecular-crystal systems.

\begin{acknowledgements}
This work was partially supported by JST FOREST (JPMJFR222G) and JSPS KAKENHI (JP23KK0088, JP23K26055), Japan. A part of the present calculations was performed on the TSUBAME4.0 supercomputer at Science Tokyo.
\end{acknowledgements}

\section*{Declaration of completing interest}
The authors declare that they have no known competing financial interests or personal relationships that could have appeared to influence the work reported in this paper.

\section*{CRediT authorship contribution statement}
Conceptualization (T.S.), Data curation (T.S, H.F.), Funding acquisition (T.S., Y.M.), Investigation (T.S. Y.M., H.F.), Methodology (T.S.), Validation (TS., H.F.), Visualization (T.S.), Writing--original draft (T.S.), Writing--review \& editing (T.S., Y.M., H.F.)

\section*{Data Availability Statement}
Data will be mede available from the corresponding authors upon reasonable request.

\bibliography{indium_iodide_phonon_Shiga}

\clearpage
\onecolumngrid

\begin{center}
  \textbf{\large Supplementary Materials for}\\[0.2cm]
  \textbf{Anharmonic lattice dynamics study of phonon transport in layered and molecular-crystal indium iodides}\\[0.3cm]

 Takuma Shiga,$^{1,\ast}$%
~Yoshikazu Mizuguchi,$^{2}$%
~ and ~Hiroshi Fujihisa$^{3}$
 \\[0.2cm]

$^{1}$\textit{Mechanical Material Engineering Laboratory, Toyota Technological Institute,
Nagoya, Aichi 468-8511, Japan}\\
$^{2}$\textit{Department of Physics, Tokyo Metropolitan University,
Hachioji, Tokyo 192-0397, Japan}\\
$^{3}$\textit{National Metrology Institute of Japan (NMIJ),
National Institute of Advanced Industrial Science and Technology (AIST),
Tsukuba, Ibaraki 305-8565, Japan}\\
{\tt shiga@toyota-ti.ac.jp}
\end{center}

\setcounter{section}{0}
\setcounter{figure}{0}
\setcounter{table}{0}
\renewcommand{\thesection}{S\arabic{section}}
\renewcommand{\theequation}{S\arabic{equation}}
\renewcommand{\thefigure}{S\arabic{figure}}
\renewcommand{\thetable}{S\arabic{table}}

\begin{table}[h]
	\caption{
	Supercell sizes used for the calculation of interatomic force constants (IFCs). $N_q$ denotes an $N_1 \times N_2 \times N_3$ mesh, where $N_i$ is the number of sampling points along the $i$th reciprocal lattice vector.
	}
	\label{tabS1}
	\centering
	\begin{tabular}{lllll}
	\hline 
	& Harmonic & Third-order & Fourth-order & $N_q$ \\
	\hline 
	\ce{InI} & \makecell[l]{$4 \times 4 \times 6$ supercell \\ (384 atoms)} & \makecell[l]{$2 \times 2 \times 3$ supercell \\ (48 atoms), \\ fifth NN (\SI{4.8575}{\angstrom}), \\ seventh NN (\SI{4.3195}{\angstrom}), \\ ninth NN (\SI{6.1124}{\angstrom})} & \makecell[l]{$2 \times 2 \times 3$ supercell \\ (48 atoms), \\ first NN (\SI{3.3714}{\angstrom})} & $16 \times 16 \times 16$\\
	\hline
	\makecell[l]{Low-pressure \ce{InI3} \\ ($P2_1/c$)} & \makecell[l]{$2 \times 4 \times 2$ supercell \\ (256 atoms)} & \makecell[l]{$1 \times 2 \times 1$ supercell \\ (32 atoms), \\ third NN (\SI{4.2628}{\angstrom}),\\ fifth NN (\SI{4.3195}{\angstrom}),\\ seventh NN (\SI{4.7302}{\angstrom})}& \makecell[l]{$1 \times 2 \times 1$ supercell \\ (32 atoms), \\ first NN (\SI{3.3564}{\angstrom})}& $8 \times 16 \times 8$\\
	\hline
	\makecell[l]{High-pressure \ce{InI3} \\ ($R\bar{3}$)} & \makecell[l]{$3 \times 3 \times 3$ supercell \\ (216 atoms)} & \makecell[l]{$2\times 2\times 2$ supercell \\ (64 atoms), \\ third NN (\SI{4.7040}{\angstrom}),\\ fifth NN (\SI{5.4436}{\angstrom}), \\ seventh NN (\SI{6.1775}{\angstrom})} & \makecell[l]{$2\times 2\times 2$ supercell \\ (64 atoms), \\ first NN (\SI{2.9651}{\angstrom})} & $16 \times 16 \times 16$\\
	\hline
	\makecell[l]{High-pressure \ce{InI3} \\ ($P\bar{3}1c$)} & \makecell[l]{$3 \times 3 \times 2$ supercell \\ (288 atoms)} & \makecell[l]{$2 \times 2 \times 1$ supercell \\ (64 atoms), \\ third NN (\SI{5.4606}{\angstrom}), \\ fifth NN (\SI{6.6830}{\angstrom}),\\ seventh NN (\SI{7.1865}{\angstrom})} & \makecell[l]{$2 \times 2 \times 1$ supercell \\ (64 atoms), \\ first NN (\SI{3.595}{\angstrom})} & $18 \times 18 \times 7$\\
	\hline
	\makecell[l]{High-pressure \ce{InI3}, \\ ($P312$)} & \makecell[l]{$3 \times 3 \times 1$ supercell \\ (216 atoms)} & \makecell[l]{$2 \times 2 \times 1$ supercell \\ (96 atoms), \\ first NN (\SI{3.5934}{\angstrom}), \\ second NN (\SI{4.7079}{\angstrom}),\\ third NN (\SI{5.4597}{\angstrom})} & \makecell[l]{$2 \times 2 \times 1$ supercell\\ (96 atoms), \\ first NN (\SI{3.5934}{\angstrom})} & $18 \times 18 \times 6$\\
	\hline
	\makecell[l]{High-pressure \ce{InI3} \\ ($P\bar{3}1m$)} & \makecell[l]{$3 \times 3 \times 3$ supercell \\ (216 atoms)} & \makecell[l]{$2 \times 2 \times 2$ supercell \\ (64 atoms), \\ third NN (\SI{5.4760}{\angstrom}), \\ fifth NN (\SI{6.6643}{\angstrom}), \\ seventh NN (\SI{7.0667}{\angstrom})} & \makecell[l]{$2 \times 2 \times 2$ supercell \\ (64 atoms), \\ first NN (\SI{3.5881}{\angstrom})} & $16 \times 16 \times 16$ \\
	\hline
	Monolayer \ce{In2I6} &  \makecell[l]{$4 \times 4 \times 1$ supercell \\ (128 atoms)} & \makecell[l]{$3 \times 3 \times 1$ supercell \\ (72 atoms), \\ first NN (\SI{3.5942}{\angstrom}),\\ second NN (\SI{4.7206}{\angstrom}),\\ third NN (\SI{5.9225}{\angstrom})} & \makecell[l]{$3 \times 3 \times 1$ supercell \\ (72 atoms), \\ first NN (\SI{3.5942}{\angstrom})} & $100 \times 100 \times 1$ \\
	\hline
	\end{tabular}
\end{table}

\begin{figure}[h]
	\centering
	\includegraphics[width=0.9\textwidth]{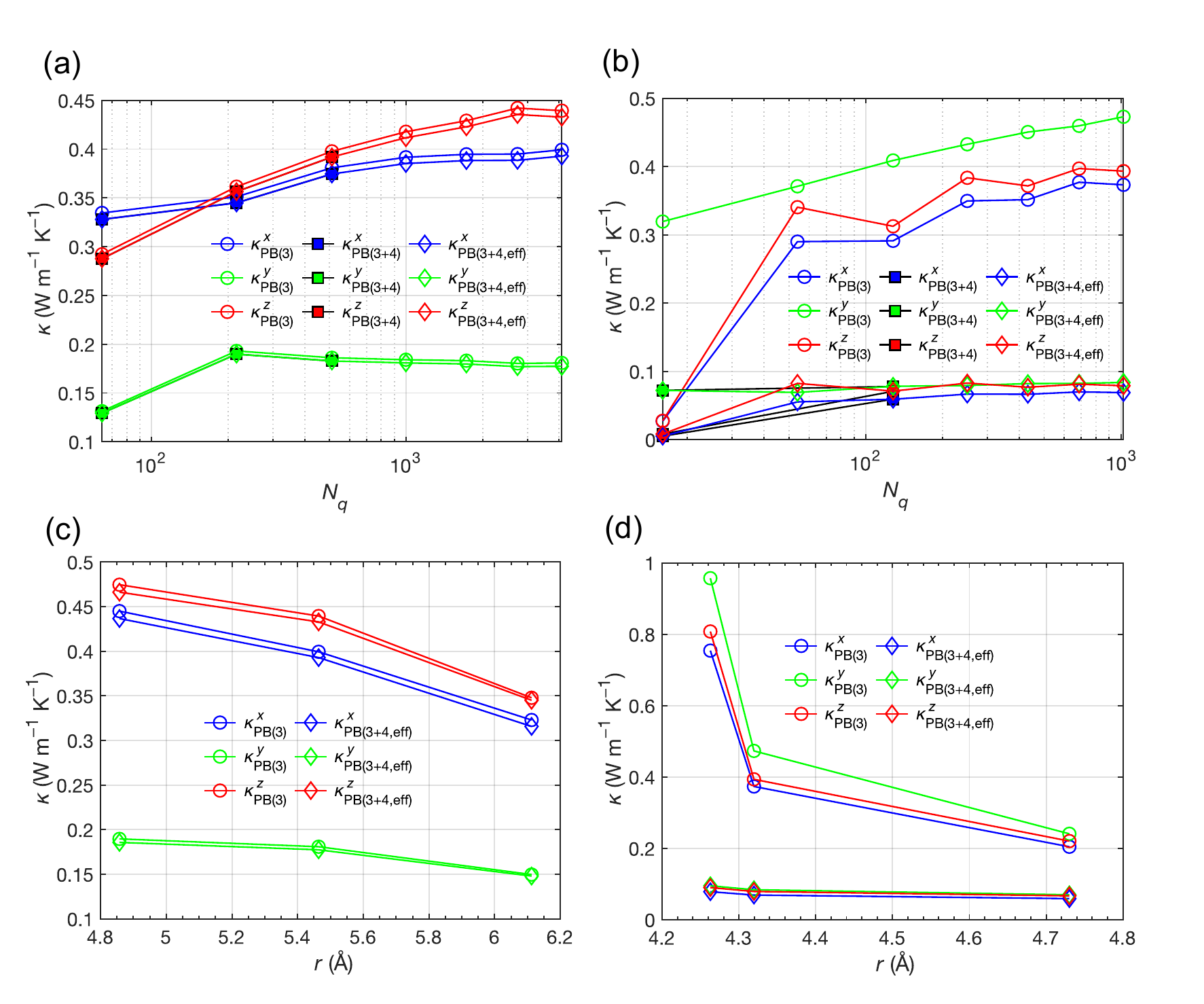}
	\caption{
	(a, b) Contribution of the particle-like Peierls component to thermal conductivity ($\kappa_{\text{PB}}$) at \SI{300}{\kelvin} along the Cartesian directions as a function of the reciprocal meshes for \ce{InI} and low-pressure \ce{InI3}, respectively. $N_q$ denotes an $N_1 \times N_2 \times N_3$ mesh, where $N_i$ is the number of sampling points along the $i$th reciprocal lattice vector. (c, d) $\kappa_{\text{PB}}$ at \SI{300}{\kelvin} as a function of the cutoff range for third-order anharmonic IFCs, calculated using $16 \times 16 \times 16$ and $8 \times 16 \times 8$ reciprocal meshes for \ce{InI} and \ce{InI3}, respectively. The open circles denote $\kappa_{\text{PB}}$ calculated with only three-phonon scattering. The filled squares and open diamonds represent $\kappa_{\text{PB}}$ with three- and four-phonon scatterings, evaluated without and with the sampling-accelerated method, respectively. Phonon--isotope scattering was included in all calculations.
	}
	\label{figS1}
\end{figure}

\begin{figure}[h]
	\centering
	\includegraphics[width=0.9\textwidth]{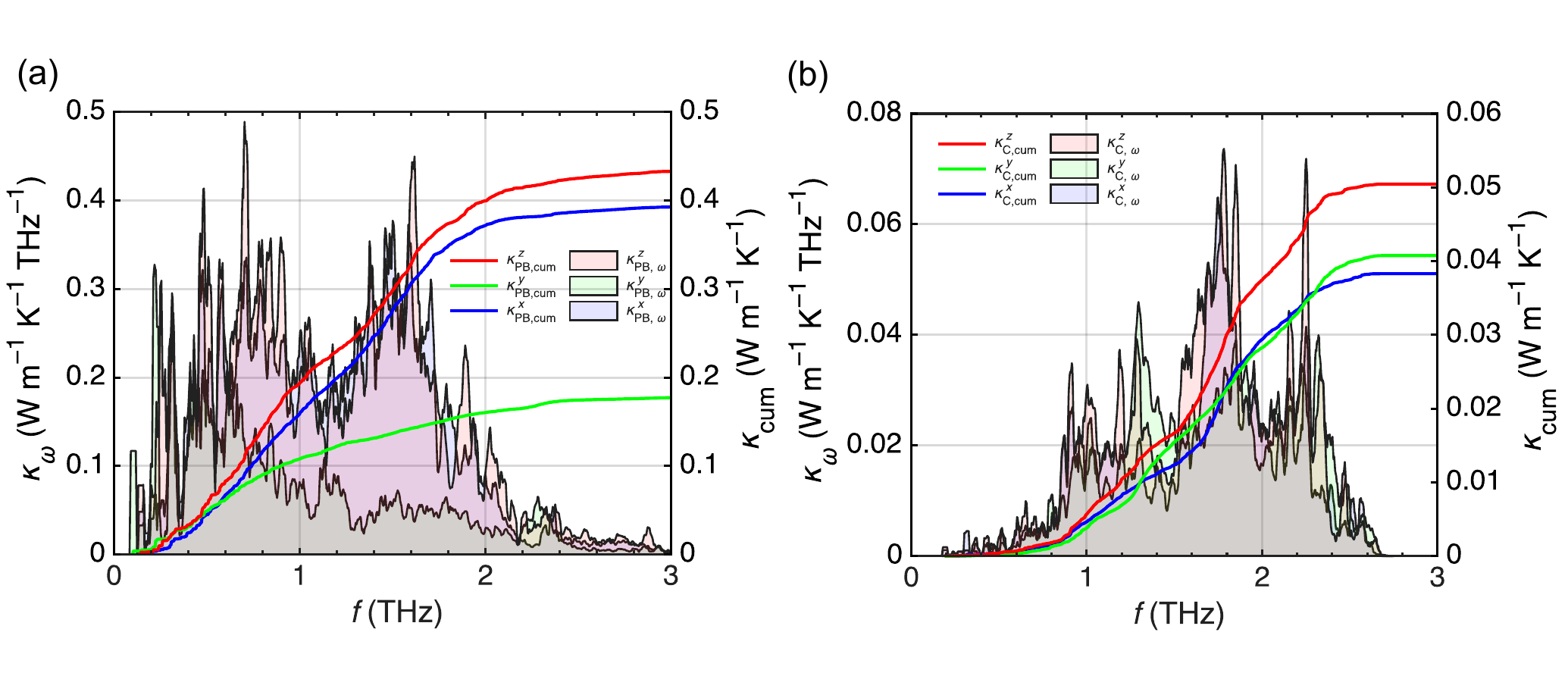}
	\caption{Spectral thermal conductivities ($\kappa_{\omega}$) of \ce{InI} at \SI{300}{\kelvin} (left axis) and their accumulations ($\kappa_{\text{cum}}$) (right axis) along the Cartesian directions: (a) Peierls ($\kappa_{\text{PB}}$) and (b) wave-like inter-band tunneling ($\kappa_{\text{C}}$) terms, respectively.}
	\label{figS2}
\end{figure}

\begin{figure}[h]
	\centering
	\includegraphics[width=0.9\textwidth]{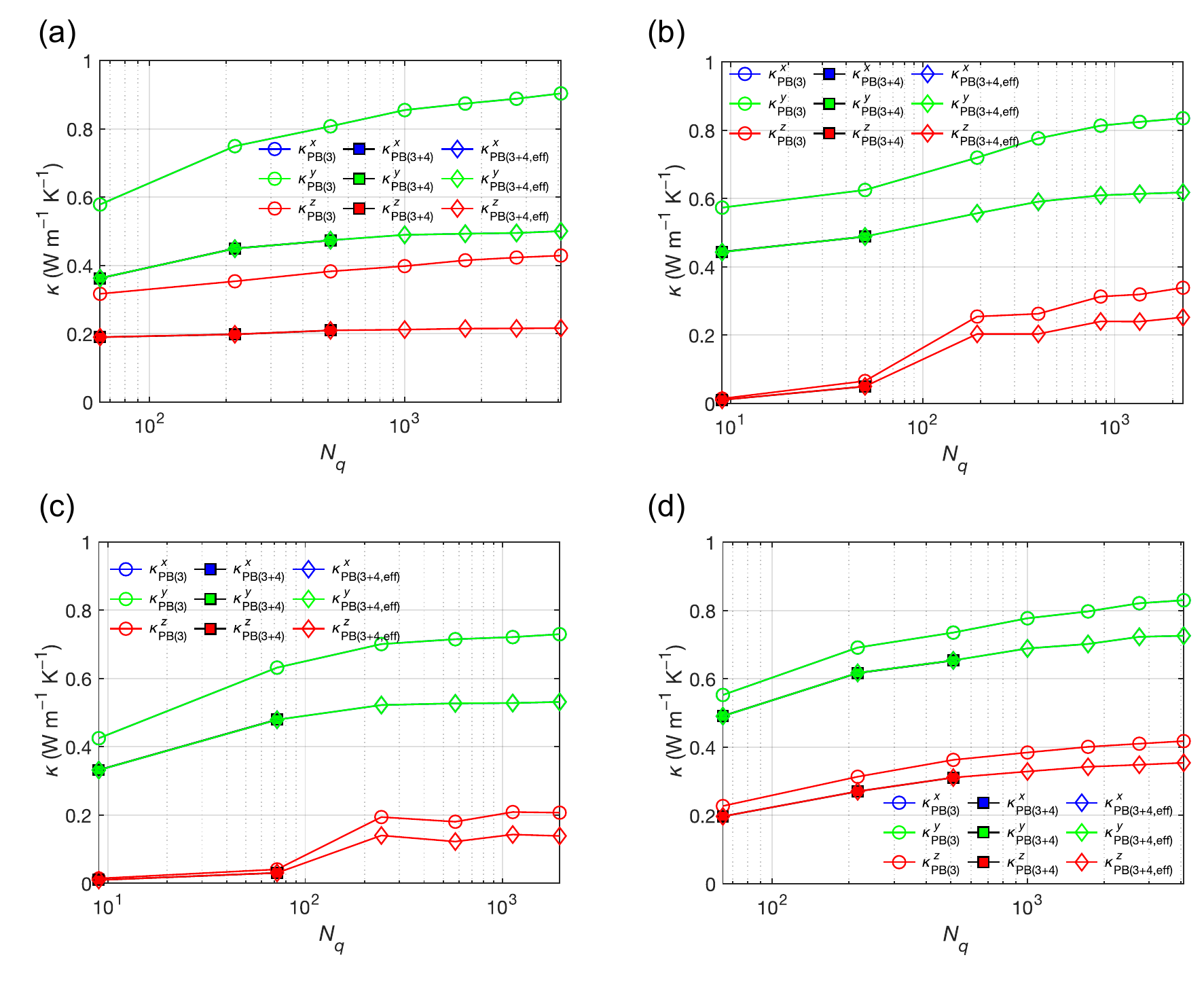}
	\caption{$\kappa_{\text{PB}}$ at \SI{300}{\kelvin} along the Cartesian directions as a function of the reciprocal mesh for the high-pressure \ce{InI3} models: (a) $R\bar{3}$, (b) $P\bar{3}1c$, (c) $P312$, and (d) $P\bar{3}1m$. The cutoff ranges for the third-order IFCs for these structural modes were set to 5.4, 6.7, 5.5, and \SI{6.7}{\angstrom}, respectively, corresponding to the fifth, fifth, third, and fifth NNs. The definition of $N_{q}$ and marker notations are the same as those in Fig. \ref{figS1}.}
	\label{figS3}
\end{figure}

\begin{figure}[h]
	\centering
	\includegraphics[width=0.9\textwidth]{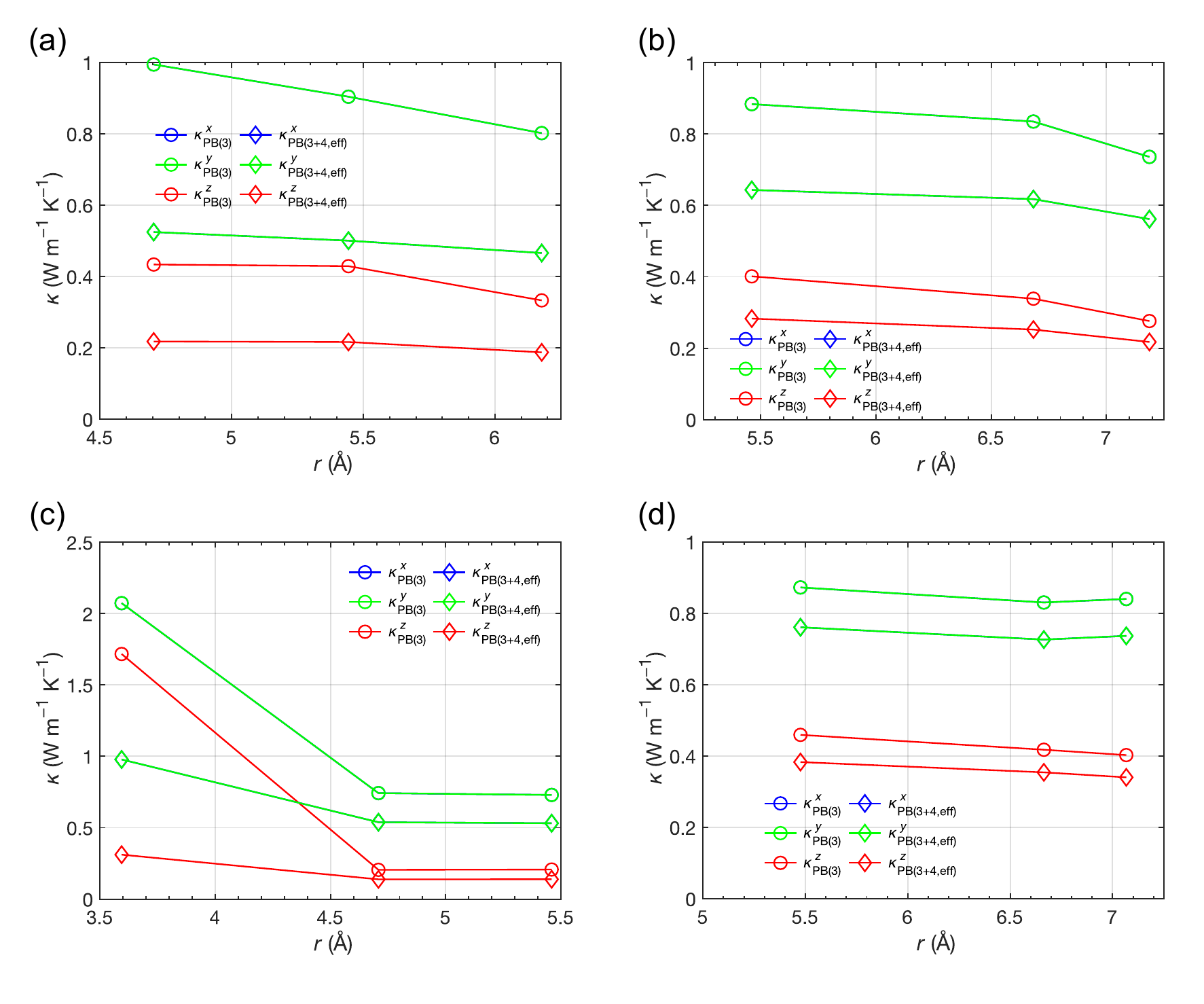}
	\caption{$\kappa_{\text{PB}}$ at \SI{300}{\kelvin} along the Cartesian directions as a function of the cutoff range for the third-order IFCs for high-pressure \ce{InI3} models: (a) $R\bar{3}$, (b) $P\bar{3}1c$, (c) $P312$, and (d) $P\bar{3}1m$. The reciprocal meshes used for these structures were $16\times 16\times 16$, $18\times 18 \times 7$, $18\times 18\times6$, and $16\times 16 \times 16$, respectively. The marker definitions are the same as those in Fig. \ref{figS1}.}
	\label{figS4}
\end{figure}

\begin{figure}[h]
	\centering
	\includegraphics[width=0.9\textwidth]{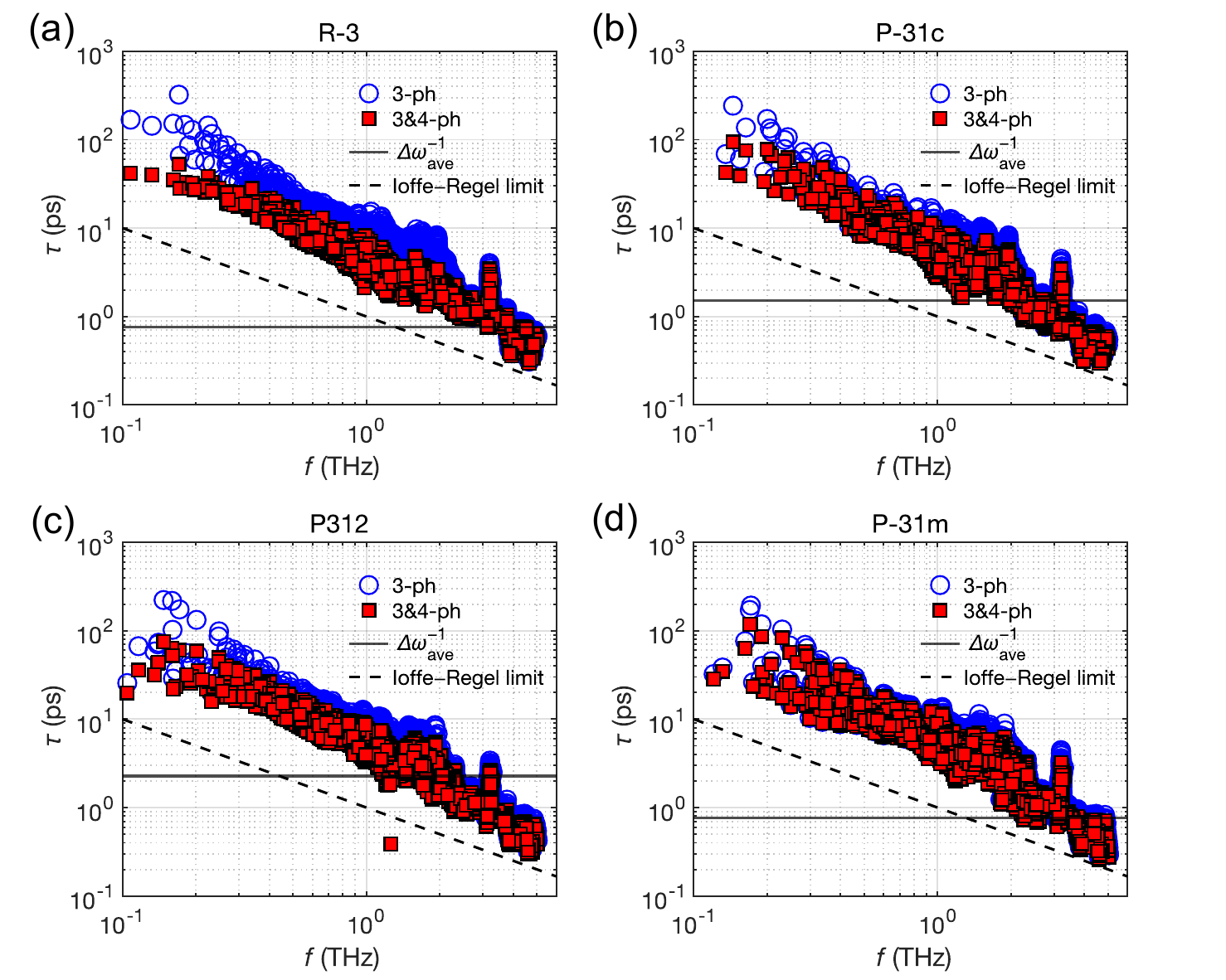}
	\caption{Frequency-dependent relaxation times ($\tau$) at \SI{300}{\kelvin} for the high-pressure \ce{InI3} models: (a) $R\bar{3}$, (b) $P\bar{3}1c$, (c) $P312$, and (d) $P\bar{3}1m$. The blue open circles and red filled squares represent $\tau$ with only three-phonon and both three- and four-phonon scatterings, respectively. The solid and dashed lines correspond to the inverse of the average interband spacing ($\Delta\omega_{\text{ave}}^{-1}$) and the Ioffe--Regel limit ($2\pi/\omega$), respectively. Phonon--isotope scattering was included in all calculations.}
	\label{figS5}
\end{figure}

\begin{figure}[h]
	\centering
	\includegraphics[width=0.9\textwidth]{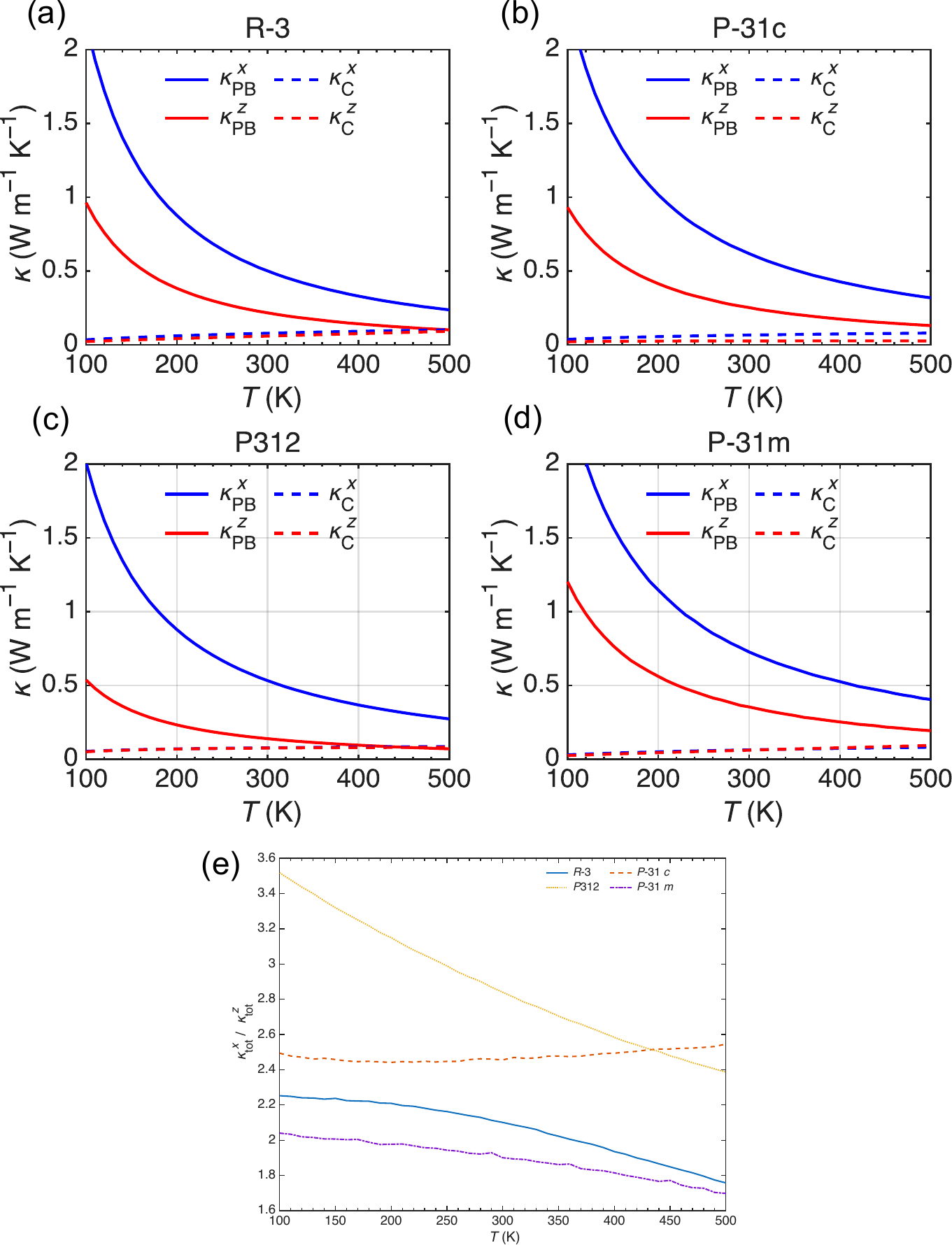}
	\caption{$\kappa_{\text{PB}}$ and $\kappa_{\text{C}}$ contributions to the temperature-dependent thermal conductivity along the Cartesian directions for high-pressure \ce{InI3} models: (a) $R\bar{3}$, (b) $P\bar{3}1c$, (c) $P312$, and (d) $P\bar{3}1m$, respectively. (e) Temperature-dependent anisotropy of the total thermal conductivity ($\kappa_{\text{tot}} = \kappa_{\text{PB}} + \kappa_{\text{C}}$), expressed as the ratio of out-of-plane to in-plane components, for the four high-pressure structural models.}
	\label{figS6}
\end{figure}

\begin{figure}[h]
	\centering
	\includegraphics[width=0.9\textwidth]{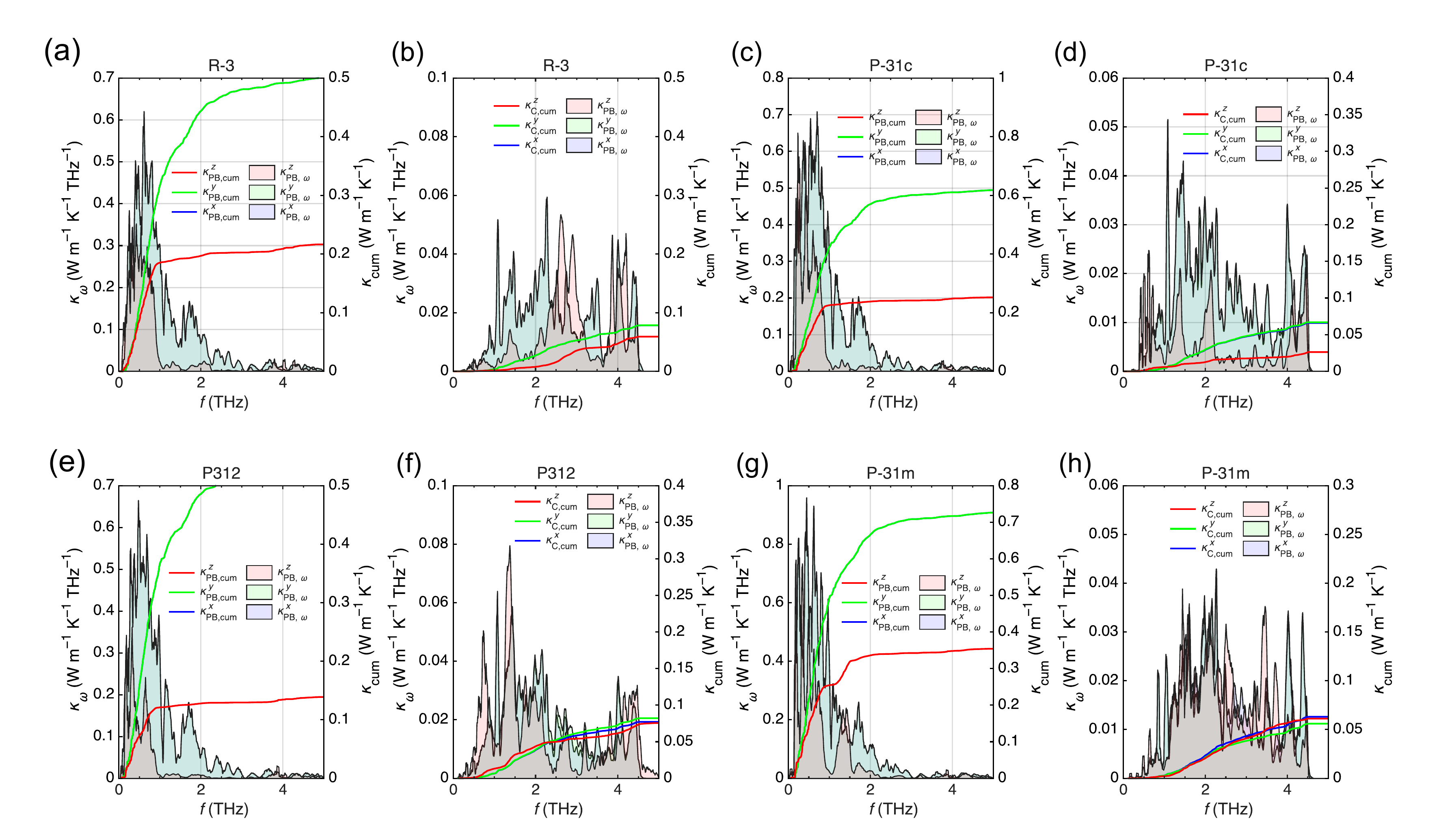}
	\caption{$\kappa_{\omega}$ at \SI{300}{\kelvin} (left axis) and their corresponding $\kappa_{\text{cum}}$ values (right axis) along the Cartesian directions for the high-pressure \ce{InI3} models: Panels (a, b), (c, d), (e, f), and (g, h) correspond to the $\kappa_{\text{PB}}$ and $\kappa_{\text{C}}$ components of the $R\bar{3}$, $P\bar{3}1c$, $P312$, and $P\bar{3}1m$ structures, respectively.}
	\label{figS7}
\end{figure}

\begin{figure}[h]
	\centering
	\includegraphics[width=0.9\textwidth]{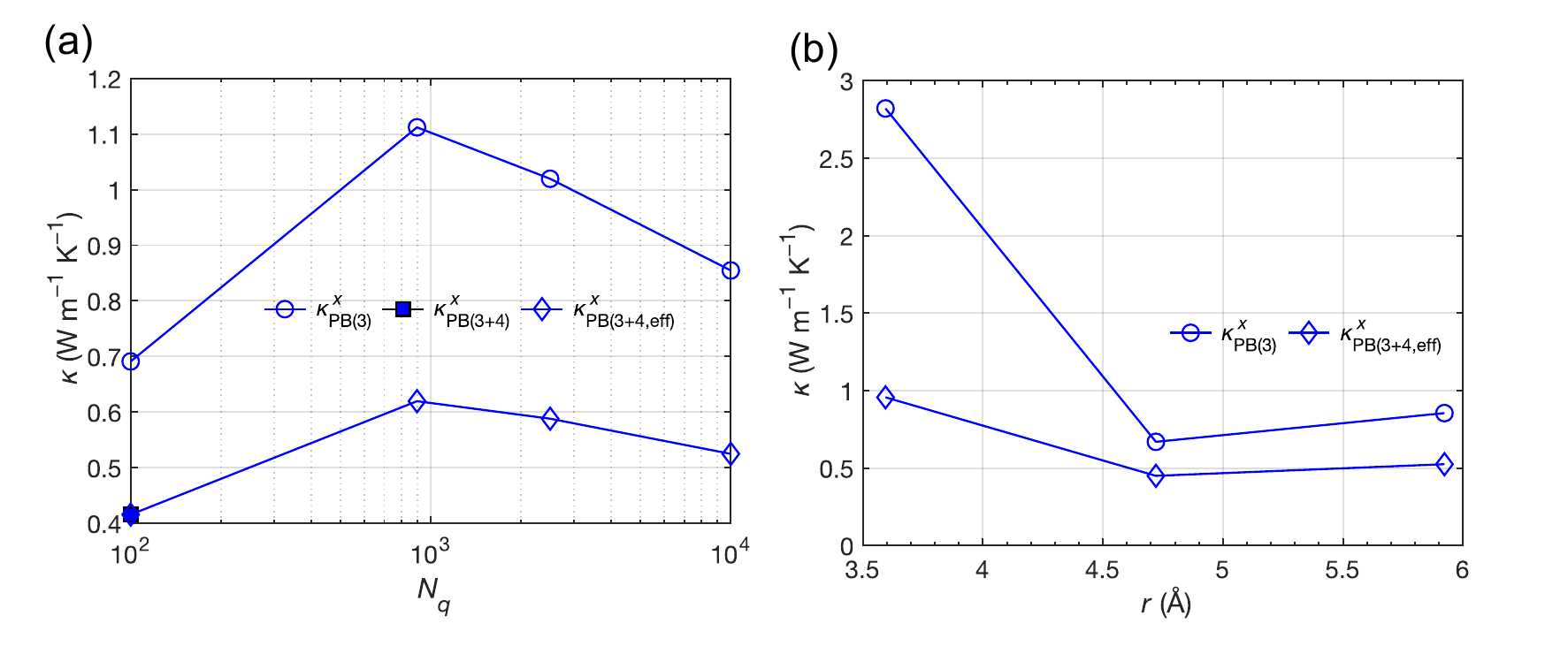}
	\caption{$\kappa_{\text{PB}}$ of monolayer \ce{In2I6} at \SI{300}{\kelvin} along the Cartesian directions: (a) reciprocal mesh and (b) cutoff ranges. The definition of $N_{q}$ and marker notations are the same as those in Fig. \ref{figS1}.}
	\label{figS8}
\end{figure}

\end{document}